\documentclass[10pt,twocolumn,letterpaper]{article}

\usepackage[final]{cvpr}      

\usepackage{graphicx}
\usepackage{amsmath}
\usepackage{amssymb}
\usepackage{booktabs}
\usepackage{algorithm2e}
\usepackage{enumitem}
\usepackage{subfloat}
\usepackage{lineno}
\usepackage[pagebackref,breaklinks,colorlinks]{hyperref}

\usepackage[capitalize]{cleveref}
\crefname{section}{Sec.}{Secs.}
\Crefname{section}{Section}{Sections}
\Crefname{table}{Table}{Tables}
\crefname{table}{Tab.}{Tabs.}

\usepackage{algorithmic}
\usepackage{multirow}
\usepackage{stfloats}

\usepackage{newfloat}
\usepackage{listings}

\begin{document}
\title{Turning Strengths into Weaknesses: A Certified Robustness Inspired \\ Attack Framework against Graph Neural Networks}

\author{
Binghui Wang\footnotemark[4 ] \footnotemark[1], Meng Pang\footnotemark[2] \footnotemark[1], and Yun Dong\footnotemark[3]\\
\footnotemark[4]\,\,{\small Department of Computer Science, Illinois Institute of Technology}\\
\footnotemark[2]\,\,{\small School of Mathematics and Computer Sciences, Nanchang University}\\
\footnotemark[3]\,\,{\small Department of Visual and Performing Arts, Education, and Sciences, Waubonsee Community College}\\
Email: \footnotemark[4]\,\,{\tt\small  bwang70@iit.edu}, \footnotemark[2]\,\,{\tt\small mengpang@ncu.edu.cn}, \footnotemark[3]\,\,{\tt\small ydong@waubonsee.edu}
\vspace{-5mm}
}

\maketitle

\pagestyle{empty} 
\thispagestyle{empty} 

\renewcommand{\thefootnote}{\fnsymbol{footnote}}
\footnotetext[1]{Corresponding authors}
\renewcommand{\thefootnote}{\arabic{footnote}}

\begin{abstract}
\label{sec:abstract}

Graph neural networks (GNNs) have achieved state-of-the-art performance in many graph learning tasks. 
However, recent studies show that GNNs are vulnerable to both test-time evasion 
and training-time  poisoning  attacks that perturb the graph structure. 
While existing attack methods have shown promising attack performance, 
we would like to design an attack framework to further enhance the  performance. 
In particular, our attack framework is inspired by certified robustness, which
was originally used by \emph{defenders} to defend against adversarial attacks. We are the first, from the \emph{attacker} perspective, to leverage its properties to better attack GNNs. 
Specifically, we first derive nodes' certified perturbation sizes against graph evasion and poisoning attacks based on randomized smoothing, respectively. A larger certified perturbation size of a node indicates this node is \emph{theoretically} more robust to graph perturbations. Such a property motivates us to focus more on nodes with smaller certified perturbation sizes, as they are easier to be attacked after graph perturbations. 
Accordingly, we design a certified robustness inspired attack loss, when incorporated into (any) existing attacks, produces our certified robustness inspired attack counterpart. 
We apply our  framework to the existing attacks and results show it can significantly enhance the existing base attacks' performance. 

\end{abstract}

\section{Introduction}
\label{sec:intro}

Learning with graphs, such as social networks, citation networks, chemical networks, has attracted significant attention recently.  Among many methods, graph neural networks (GNNs)~\cite{kipf2017semi,wu2019simplifying,velivckovic2018graph,xu2019powerful,zhang2018link} have achieved state-of-the-art performance in graph related tasks such as node classification, graph classification, and link prediction.  
However, recent studies~\cite{zugner2018adversarial,dai2018adversarial,zugner2019adversarial,xu2019topology,liu2019unified,sun2020adversarial,wu2019adversarial,wang2019attacking,ma2020towards,mu2021hard,wang2020evasion,wang2022bandits} show that GNNs are vulnerable to both test-time graph evasion attacks and training-time graph poisoning attacks\footnote{We mainly consider the graph structure attack in the paper, as it is more effective than the feature attack. 
However, our attack framework can be easily extended to the feature attack.}.
Take GNNs for node classification as an instance, graph evasion attacks mean that, given a learnt GNN model and a (clean) graph, an attacker carefully perturbs the graph structure (i.e., inject new edges to or remove  the existing edges from the graph) such that as many testing nodes as possible are misclassified 
by the GNN model. 
Whereas, graph poisoning attacks mean that, given a GNN algorithm and a graph, an attacker carefully perturbs the graph structure in the training phase, such that the learnt GNN model misclassifies as many testing nodes as possible 
in the testing phase. 
While existing methods have shown promising attack performance, 
we want to ask: 
Can we design a general \emph{attack framework}
that can further enhance both the existing graph evasion and poisoning attacks to GNNs? The answer is yes. 

We design an attack framework inspired by certified robustness.
Certified robustness was originally used by \emph{defenders} to guarantee the robustness of classification
models against evasion attacks. 
Generally speaking, a testing example (e.g., an image or a node) with a better certified robustness guarantee indicates this example is \emph{theoretically}  more robust to adversarial (e.g., pixel or graph) perturbations. 
While certified robustness is mainly derived for doing the good, \emph{attackers}, on the other hand, can also leverage its property to do the bad.
For instance, when an attacker knows the certified robustness of nodes in a graph, he can base on nodes' certified robustness to 
\emph{reversely} reveal the vulnerable region of the graph and leverage this vulnerability to design better attacks. 
We are inspired by such property of certified robustness and 
design the first certified robustness inspired attacks to GNNs.

Our attack framework consists of three parts: 
i) Inspired by the state-of-the-art randomized smoothing based certified robustness against \emph{evasion attacks} to image models~\cite{cohen2019certified,salman2019provably} and GNN models~\cite{wang2021certified}, we first propose to {generalize} randomized smoothing and derive the {node's certified perturbation size} against graph \emph{poisoning attacks} to GNNs.
 Particularly, a larger certified perturbation size of a node indicates  
 this node is \emph{theoretically} more robust to adversarial graph perturbations. In other words, an attacker needs to perturb more edges during the training phase in order to make this node wrongly predicted by the learnt GNN model. This property inspires us to focus more on disrupting nodes with relatively smaller certified perturbation sizes under a given perturbation budget. 
 ii) We design a certified robustness inspired attack loss. Specifically,  we modify the classic node-wise loss 
 by assigning each node a weight based on its certified perturbation size---A node with a larger/smaller certified perturbation size will be assigned a smaller/larger weight. 
In doing so, losses for nodes with smaller certified perturbation sizes will be enlarged, and most of the perturbation budget will be automatically allocated to perturb these nodes. Thus, more nodes will be misclassified 
with the given perturbation budget. 
iii) We design the certified robustness inspired attack framework to generate adversarial graph perturbations to GNNs, based on our certified robustness inspired attack loss. 
We emphasize that, as our new attack loss only modifies the existing attack loss with certified perturbation size defined node weights, any existing graph evasion or poisoning attack method can be used as the base attack in our framework.  

We apply our certified robustness inspired attack framework to the state-of-the-art graph evasion and poisoning attacks~\cite{xu2019topology,zugner2019adversarial} to GNNs.
Evaluation results on multiple benchmark  datasets show our attack framework can substantially enhance the attack performance  
of the base attacks.
Our contributions are as follows:

\begin{itemize}[leftmargin=*]
\vspace{-2mm}
    \item We propose a certified robustness inspired attack framework to GNNs. Our framework can be 
    plugged into any existing graph evasion and poisoning attacks.
    
\vspace{-2mm}
    \item To our best knowledge, we are the first work to use certified robustness for an attack purpose.

\vspace{-2mm}    
    \item Evaluation results validate the effectiveness of our attack framework when applied to the existing attacks to GNNs. 
    
\end{itemize}

\section{Background and Preliminaries}
\label{sec:background}

\subsection{Graph Neural Networks (GNNs)}
 Let $G=(\mathcal{V}, \mathcal{E})$ be a graph, where $u \in \mathcal{V}$ is a node, $(u, v) \in \mathcal{E}$ is an edge between $u$ and $v$.
Let $\mathbf{A} \in \{0,1\}^{|\mathcal{V}| \times |\mathcal{V}|}$ 
be the adjacency matrix. 
\emph{As $\mathbf{A}$ contains all graph structure information, we will interchangeably use $\mathbf{A}$ and $G$ to indicate the graph in the paper.} 
We mainly consider GNNs for node classification.
Each node $u \in \mathcal{V}$ has a label $y_u$
 from a label set $\mathcal{Y}$. 
Let $\mathcal{V}_{Tr}$ and $\mathcal{V}_{Te}$ be the set of training nodes and testing nodes, respectively.  
Given a GNN algorithm $\mathcal{A}$, which takes the graph $G(\mathbf{A})$ and training nodes $\mathcal{V}_{Tr}$ as an input and produces a node classifier  $f_\theta$
parameterized by $\theta$, i.e., $f_\theta =\mathcal{A}(\mathbf{A}, \mathcal{V}_{Tr})$. The node classifier $f_\theta$ inputs 
$G(\mathbf{A})$ and outputs labels for  all nodes, i.e., $f_\theta: \mathbf{A} \rightarrow \mathcal{Y}^{|\mathcal{V}|}$.
To learn $f_\theta$, a common way is to minimize a loss
function $\mathcal{L}$ defined on the training nodes $\mathcal{V}_{Tr} $ and the graph $G(\mathbf{A})$ as follows:
\begin{small}
\begin{align}
    {\min _\theta }  \mathcal{L}(f_\theta, \mathbf{A}, \mathcal{V}_{Tr}) =  \sum_{\small u \in \mathcal{V}_{Tr}} \ell(f_\theta(\mathbf{A};u), y_u),
    \label{equ:gnn}
\end{align}
\end{small}%
where $f_\theta(\mathbf{A};u)$ is the predicted label of a node $u$. 
After learning $f_{\theta^*}$, a testing node $v \in \mathcal{V}_{Te}$ is then predicted a label as 
$\hat{y}_v = f_{\theta^*}(\mathbf{A};v)$.

\subsection{Adversarial Attacks to GNNs}
We denote by $\delta \in \{0,1\}^{|\mathcal{V}| \times |\mathcal{V}|}$ the adversarial \emph{graph perturbation}, where $\delta_{s,t}=1$ (or $0$) means the attacker perturbs (or keeps) the edge status between a node pair $(s,t)$. 
Moreover, we denote $\mathbf{A} \oplus \delta$ as the perturbed graph,
with $\oplus$ the element-wise XOR operator. 
For instance, if there is an (or no) edge between $(u,v)$, i.e., $A_{uv}=1$ (or $A_{uv}=0$), perturbing this edge status (i.e, $\delta_{u, v}=1$) means removing the edge (or injecting a new edge), i.e., $A_{u,v} \oplus \delta_{u,v} = 0$ (or $A_{u,v} \oplus \delta_{u,v} = 1$) to the graph. 
We assume an attacker has a perturbation budget $\Delta$, i.e., $\|\delta\|_0 \leq \Delta$, meaning at most $\Delta$ number of edges can be perturbed by the attacker. 

\noindent {\bf Graph evasion attacks to GNNs.}
In graph evasion attacks, given a learnt node classifier $f_{\theta^*}$, an attacker carefully crafts a graph perturbation $\delta$ to the graph $G$ such that $f_{\theta^*}$
predicts nodes' labels using the perturbed graph $\mathbf{A} \oplus\delta$ 
as the attacker desires. 
For instance, an attacker desires as many testing nodes as possible to be misclassified by $f_{\theta^*}$ (called \emph{untargeted attack}) 
under the perturbation budget $\Delta$. Formally, an attacker aims to maximize the following 0-1 \emph{(attack) loss}: 
\begin{small}
\begin{align}
\label{untargeted_evasion_raw}
& \max_{\delta} \sum_{v \in \mathcal{V}_{Te}} \mathbf{1}[{ f_{\theta^*}(\mathbf{A} \oplus \delta; v) \neq y_v}], 
\textrm{s.t. } ||\delta||_0 \le \Delta,
\end{align}
\end{small}%
where $\mathbf{1}[\cdot]$ is an indicator function, whose value is 1 if the condition satisfies and 0, otherwise.

The above problem is challenging to solve in that the indicator function is hard to be optimized. In practice, an attacker will solve an alternative optimize problem 
as below: 
\begin{small}
\begin{align}
& \max_{\delta} \sum_{v \in \mathcal{V}_{Te}} \ell(f_{\theta^*}(\mathbf{A} \oplus \delta; v),y_v), \, 
\textrm{s.t. } ||\delta||_0 \le \Delta. 
\label{evasion_constrain}
\end{align}
\end{small}%
For instance, \cite{xu2019topology} design the state-of-the-art PGD evasion attack by solving Equation~\ref{evasion_constrain}. 

\noindent {\bf Graph poisoning attacks to GNNs.}
In graph poisoning attacks, an attacker specifies a GNN algorithm $\mathcal{A}$ and carefully perturbs the graph $G$ with a  graph perturbation $\delta$ in the training phase, such that the learnt node classifier $f_{\theta^*}$ misclassifies as many testing nodes as possible on the perturbed graph $\mathbf{A} \oplus \delta$ in the testing phase. Formally, it solves the following bilevel optimization problem:
{
\small
\begin{align}
& \max_{\delta} \sum_{v \in \mathcal{V}_{Te}} \mathbf{1}[{ f_{\theta^*}(\mathbf{A} \oplus \delta; v) \neq y_v}], \,  \label{poison_indicator} 
\\ 
& \textrm{s.t. } \theta^* = \arg\min_\theta \sum_{\small u \in \mathcal{V}_{Tr}} \mathbf{1}[{ f_{\theta^*}(\mathbf{A} \oplus \delta; u) \neq y_u}], 
\, ||\delta||_0 \le \Delta, \nonumber 
\end{align}
}%
where the inner optimization problem is learning the node classifier $f_{\theta^*}$ on the perturbed graph $\mathbf{A} \oplus \delta$ with training nodes $\mathcal{V}_{Tr}$, while the outer optimization problem is learning to generate the graph perturbation $\delta$ to maximally misclassify testing nodes $\mathcal{V}_{Te}$ with the learnt node classifier $f_{\theta^*}$. 

In practice, the testing nodes $\mathcal{V}_{Te}$'s labels are unavailable during training, and thus we cannot directly optimize 
Equation~\ref{poison_indicator}. 
In addition, the indicator function in  Equation~\ref{poison_indicator} is hard to optimize. 
A common strategy to address this issue is by instead maximizing the loss on the \emph{training nodes} $\mathcal{V}_{Tr}$~\cite{zugner2019adversarial,xu2019topology} and using an alternative continuous loss. Specifically, it solves the following alternative bilevel optimization problem: 
\begin{small}
\begin{align}
& \max_{\delta} \sum_{v \in \mathcal{V}_{Tr}} \ell({ f_{\theta^*}(\mathbf{A} \oplus \delta; v), y_v}), \, \label{graphpoison_replace} \\ 
& \textrm{s.t. } \theta^* = \arg\min_\theta \sum_{\small u \in \mathcal{V}_{Tr}} \ell(f_\theta(\mathbf{A} \oplus \delta; u), y_u), \, ||\delta||_0 \le \Delta. \nonumber 
\end{align}
\end{small}%
This is based on the intuition that if a node classifier misclassifies a large number of training nodes, then it generalizes poorly and thus is also very likely to misclassify a large number of testing nodes. 

\subsection{Certified Robustness to Graph Evasion Attacks}
\label{bg_evasion_cr}

We introduce certified robustness achieved via the state-of-the-art randomized smoothing~\cite{lecuyer2018certified,li2018second,cohen2019certified}. 
{Randomized smoothing} was originally designed to build certified robust machine learning classifiers against evasion attacks.
It is applicable to any classifier and scalable to large models, e.g., deep neural networks.
Here, we introduce randomized smoothing that defends against graph evasion attacks to GNNs~\cite{wang2021certified}. 
It consists of the following three steps. 

\noindent {\bf Constructing a smoothed node classifier.}
Given a base node classifier $f$, a graph $G$, and a testing node $u$ with label $y_u$, randomized smoothing builds a \emph{smoothed node classifier} $g$ via adding a random noise matrix $\epsilon$ to 
$G$. Formally, 
\begin{align}
\label{smoothclassifier}
    g(\mathbf{A};u) = \arg\max_{c \in \mathcal{Y}}\text{Pr}(f(\mathbf{A}\oplus\epsilon; u)=c),
\end{align}
where $\text{Pr}(f(\mathbf{A}\oplus\epsilon; u)=c)$ is the probability that the base node classifier $f$ predicts label $c$ on the noisy graph $
\mathbf{A} \oplus \epsilon$ and $g(\mathbf{A};u)$ is the predicted label for $u$ by the smoothed node classifier $g$. 
$\epsilon$ has the following probability distribution in the binary space $\{0,1\}^{|\mathcal{V}| \times |\mathcal{V}|}$:
\begin{align}
\label{discretenoisedistribution}
\text{Pr}(\epsilon_{s,t}=0) =\beta,\ \text{Pr}(\epsilon_{s,t}=1) =1-\beta, \, \forall s,t \in \mathcal{V}. 
\end{align}
Equation~\ref{discretenoisedistribution} means that for each pair of nodes $(s,t)$ in the graph, we keep its edge status (i.e., $A_{s,t}$) with probability $\beta$ and change its edge status with probability $1-\beta$. 

\noindent {\bf Deriving the certified robustness of graph evasion attacks to GNNs.} Suppose $g(\mathbf{A};u)=y_u$, meaning that the smoothed node classifier $g$ correctly predicts $u$. Then, 
$g$ provably predicts the correct label for $u$ once the graph perturbation $\delta$ is bounded. Formally
~\cite{wang2021certified}:
\begin{align}
\label{certifiedradius}
    g(\mathbf{A} \oplus\delta; u) = y_u, \forall ||\delta||_0 \leq K(\underline{p_{y_u}}), 
\end{align}
where $\underline{p_{y_u}} \leq \textrm{Pr}(f(\mathbf{A}\oplus\epsilon; u)=y_u)$ is a lower bound of the probability that $f$ predicts the correct label $y_u$ on the noisy graph $\mathbf{A}\oplus\epsilon$. 
$K(\underline{p_{y_u}})$ is called node $u$'s \emph{certified perturbation size}, indicating that $g$ provably 
predicts the correct label when an attacker \emph{arbitrarily} perturbs 
(at most) $K(\underline{p_{y_u}})$ edge status in the graph $G$. \emph{In other words, if a node has a larger certified perturbation size, then it is certifiably more robust to adversarial graph perturbation.}

\noindent {\bf Computing the certified perturbation size in practice.} 
Note that $K(\underline{p_{y_u}})$ is (positively) related to 
$\underline{p_{y_u}}$, which can be estimated via the Monte Carlo algorithm~\cite{cohen2019certified,wang2021certified}. 
Specifically, given a node classifier $f$, a graph $G (\mathbf{A})$, and a testing node $u$, we 
first sample $N$ random noise matrices $\epsilon^1, \cdots,\epsilon^N$ from the noise distribution defined in Equation~\ref{discretenoisedistribution} and add each noise matrix $\epsilon^j$ to the graph $G$ to construct $N$ noisy graphs $\mathbf{A}\oplus\epsilon^1, \cdots, \mathbf{A}\oplus\epsilon^N$.  
Then, we use the node classifier $f$ to predict $u$'s label on the $N$ noisy graphs and compute the frequency of each label $c$, i.e., $N_c=\sum_{j=1}^{N}\mathbb{I}({f}(\mathbf{A}\oplus\epsilon^j, , u)=c)$ for $c\in \mathcal{Y}$.
Then, we can estimate $\underline{p_{y_u}}$ as 
\begin{align}
\label{eqn:lowbound}
\underline{p_{y_u}} = B(\alpha; N_{y_u}, N-N_{y_u}+1),
\end{align}
where $1 -\alpha$ is the confidence level and $B(\alpha; a, b)$ is the $\alpha$-th quantile of Beta distribution with shape parameters $a$ and $b$.
With $\underline{p_{y_u}}$, we can compute $K(\underline{p_{y_u}})$, and details of computing $K(\underline{p_{y_u}})$  can been seen in \cite{wang2021certified}.

\section{Certified Robustness to Graph Poisoning Attacks via Randomized Smoothing}
\label{bg_poison_cr}

Existing randomized smoothing mainly certifies robustness of \emph{evasion attacks}. 
In this section, 
we generalize it 
and derive certified robustness of graph poisoning attacks. 
Our key idea is to extend randomized smoothing from the \emph{classifier} perspective to a general \emph{function} perspective. In particular, we will 
build a base function, a smoothed function, and then  adapt randomized smoothing to certify robustness to poisoning attacks using the smoothed function.
Such certified robustness guides us to design more effective graph poisoning attacks, as shown in Section~\ref{sec:cr_attack}. 

\noindent {\bf Building a base function.} Suppose we have a graph $G(\mathbf{A})$, training nodes $\mathcal{V}_{Tr}$, and a GNN algorithm $\mathcal{A}$ that takes the graph and training nodes as an input and learns a node classifier $f$, i.e., $f=\mathcal{A}(\mathbf{A}, \mathcal{V}_{Tr})$.  
We use the learnt 
$f$ to predict the label for a testing node $v$. 
Then, we can integrate the entire process of training the node classifier $f$ and testing the node $v$ as a 
function $\tilde{f}(\mathbf{A}, \mathcal{V}_{Tr}; v)$. 
In other words, the function $\tilde{f}$ is the composition of learning the node classifier $f$ and predicting a node $v$.
We view $\tilde{f}$ as the base function. 

\noindent {\bf  Constructing a smoothed function.}  
In graph poisoning attacks, an attacker aims to perturb the graph in the training phase. To apply randomized smoothing, we first add a random noise matrix $\epsilon$ to the graph, where each entry $\epsilon_{s,t}$ is drawn from a discrete distribution, e.g.,  defined in Equation~\ref{discretenoisedistribution}. 
As we add random noise $\epsilon$ to the graph $G$, the
output of the base function $\tilde{f}$ is also random. 
Then, inspired by Equation~\ref{smoothclassifier}, we define the smoothed function $\tilde{g}$ as follows: 
\begin{align}
\label{smoothclassifier_poison}
    \tilde{g}(\mathbf{A},\mathcal{V}_{Tr}; v) = \arg\max_{c \in \mathcal{Y}}\text{Pr}(\tilde{f}(\mathbf{A}\oplus\epsilon, \mathcal{V}_{Tr}; v)=c),
\end{align}
where $\textrm{Pr}(\tilde{f}(\mathbf{A}\oplus\epsilon,\mathcal{V}_{Tr};v)=c)$ is the probability that $v$ is predicted to be a label $c$ by a GNN model trained on a noisy graph $\mathbf{A}\oplus \epsilon$ using training nodes $\mathcal{V}_{Tr}$. $\tilde{g}(\mathbf{A},\mathcal{V}_{Tr}; v) $ is the predicted label for $v$ by 
the smoothed function $\tilde{g}$. 

\noindent {\bf  Deriving the certified robustness of graph poisoning attacks to GNNs.} An attacker adds an adversarial graph perturbation $\delta$ to the graph $G(\mathbf{A})$ to produce a perturbed graph $\mathbf{A} \oplus \delta$, where $\delta_{s,t}$ is the perturbation added to change the edge status of 
the node pair $(s,t)$ in the graph $G$ during training. 
Then, we can leverage the results in Equation~\ref{certifiedradius} to derive the certified perturbation size against graph poisoning attacks. 
Specifically, we have: 
\begin{align}
\label{certify_poison}
\tilde{g}(\mathbf{A} \oplus \delta, \mathcal{V}_{Tr}; v) = y_v, \ \forall ||\delta||_0 \leq K(\underline{p_{y_v}}),
\end{align}
where $\underline{p_{y_v}} \leq \textrm{Pr}(\tilde{f}(\mathbf{A}\oplus\epsilon,\mathcal{V}_{Tr};v)=y_v)$ is a lower bound probability. 
Our result means 
the smoothed function $\tilde{g}$ provably predicts the correct label for $v$ when (at most) $K(\underline{p_{y_v}})$ edge statuses
in the graph are \emph{arbitrarily} poisoned by an attacker \emph{in the training phase}. 

\noindent {\bf  Computing the certified perturbation size in practice.} 
Given a GNN algorithm $\mathcal{A}$, a graph $G(\mathbf{A})$, training nodes $\mathcal{V}_{Tr}$, a discrete noise distribution defined in 
Equation~\ref{discretenoisedistribution}, and a node $v$,
we first sample $N$ random noise matrices $\epsilon^1, \cdots,\epsilon^N$ from the discrete noise distribution and add each noise to the graph $G(\mathbf{A})$ to construct $N$ noisy graphs $\mathbf{A}\oplus\epsilon^1,\cdots, \mathbf{A}\oplus\epsilon^N$.  
Then, we train $N$ node classifiers $\tilde{f}^1=\mathcal{A}(\mathbf{A}\oplus\epsilon^1, \mathcal{V}_{Tr}), \cdots, \tilde{f}^N=\mathcal{A}(\mathbf{A}\oplus\epsilon^N, \mathcal{V}_{Tr})$. We use each of the $N$ node classifiers to predict $v$'s label and compute the frequency of each label $c$, i.e., $N_c=\sum_{j=1}^{N}\mathbb{I}(\tilde{f}^j(\mathbf{A}\oplus\epsilon^j, \mathcal{V}_{Tr}; v)=c)$ for $c\in \mathcal{Y}$. Finally, we estimate $\underline{p_{y_v}}$ using Equation~\ref{eqn:lowbound} and use it to calculate the certified  perturbation size, following~\cite{wang2021certified}. 
Note the trained $N$ node classifiers 
is re-used to predict node labels and compute certified  perturbation size  for different nodes.

\section{Certified Robustness Inspired Attack Framework against GNNs}
\label{sec:cr_attack}
In this section, we will design our attack framework to GNNs 
inspired by certified robustness. Our attack framework can be seamlessly plugged into the existing graph evasion and poisoning attacks to design more effective attacks.

\subsection{Motivation and Observation} 
Certified robustness, more specifically, certified perturbation size derived in Section~\ref{bg_evasion_cr} and Section~\ref{bg_poison_cr}, was 
used by \emph{defenders} 
to {defend} GNN models against attacks.  
On the other hand, from the \emph{attacker} perspective, 
he can leverage the properties of certified robustness to 
better attack GNN models. 
Specifically, 
certified perturbation size of a node characterizes the extent to which the GNN model \emph{provably} and accurately predicts this node against the worst-case graph perturbation. 
An attacker can base on nodes' certified perturbation sizes to 
\emph{reversely} reveal the vulnerable region of the graph and leverage this vulnerability to design better attacks. 
In particular, we have the following observation that reveals the \emph{inverse} relationship between a node's certified perturbation size and the perturbation associated with this node when designing the attack.

\vspace{+0.5mm}
\emph{\textbf{Observation 1:} A node with a larger (smaller) certified perturbation size should be disrupted with a smaller (larger) number of perturbed edges.}
\vspace{+0.5mm}

If a node has a larger (smaller) certified perturbation size, it means this node is more (less) robust to graph perturbations. To misclassify this node, an attacker should allocate a larger (smaller) number of perturbed edges. 
Thus, to design more effective attacks (i.e., misclassify more nodes) with a perturbation budget, an attacker should avoid disrupting nodes with larger certified perturbation sizes, but focus on 
nodes with smaller certified perturbation sizes.

Based on the above observation, our attack needs to solve three correlated problems: i) How to obtain the node's certified perturbation size for both graph evasion and poisoning attacks? ii) How to allocate the perturbation budget in order to disrupt the nodes with smaller certified perturbation sizes? iii) How to generate the adversarial graph perturbation for both evasion and poisoning attacks? 
To address i), we adopt the derived node's certified perturbation size against graph evasion attacks (Section~\ref{bg_evasion_cr}) and graph poisoning attacks (Section~\ref{bg_poison_cr}).
To address ii),  we design a certified robustness inspired loss, by maximizing which an attacker will put more effort into disrupting nodes with smaller certified perturbation sizes.   
To address iii), we design a certified robustness inspired attack framework, where any existing graph evasion/poisoning attacks to GNNs can be adopted as the base attack in our framework.

\subsection{Certified Robustness Inspired Loss Design} 

Suppose we have obtained nodes' certified perturbation sizes. 
To perform a more effective attack, a naive solution is that 
the attacker sorts all nodes' certified perturbation sizes in an ascending order, and then carefully 
perturbs the edges to misclassify the sorted nodes one-by-one 
until reaching the perturbation budget. 
However, this solution is both computationally intensive---as it needs to solve an optimization problem for each node; and suboptimal---as all nodes and the associated edges collectively make predictions and perturbing an edge could affect predictions of many nodes.   
 
 We design a certified perturbation size inspired loss that assists to \emph{automatically} seeking the ``ideal" edges to be perturbed for both evasion attacks and poisoning attacks.
Particularly, we notice that the loss function of evasion attacks in Equation~\ref{evasion_constrain} or poisoning attacks in Equation~\ref{graphpoison_replace} 
 is defined per node. 
 Then, we propose to modify the loss function in Equation~\ref{evasion_constrain} or Equation~\ref{graphpoison_replace} 
 by assigning each node with a weight and multiplying each node loss with the corresponding weight, where the node weight has a strong connection with 
 the node's certified perturbation size. 
 Formally, we design the certified perturbation size inspired loss as follows: 
\begin{align}
\label{eq:cr_loss}
    \mathcal{L}_{CR}(f_\theta, \mathbf{A},\mathcal{V}_{T}) = \sum_{\small u \in \mathcal{V}_{T}} w(u) \cdot \ell(f_\theta(\mathbf{A}; u), y_u),
\end{align}
where $\mathcal{V}_{T}=\mathcal{V}_{Te}$ for evasion attacks and $\mathcal{V}_{T}=\mathcal{V}_{Tr}$ for poisoning attacks; and 
$w(u)$ is the weight of the node $u$. Note that when setting all nodes with an equal weight, our certified perturbation size inspired loss reduces to the conventional loss in Equation~\ref{evasion_constrain} or Equation~\ref{graphpoison_replace}.  
Next, we show the \emph{inverse} relationship between the
node's certified perturbation size and the assigned  weight.

\vspace{+0.5mm}
\emph{\textbf{Observation 2:} A node with a larger (smaller) certified perturbation size is 
assigned a smaller (larger) weight.} 
\vspace{+0.5mm}

As shown in {\bf Observation 1}, we should disrupt more nodes with smaller certified perturbation sizes, as these nodes are more vulnerable. In other words, we should put more weights on nodes with smaller certified perturbation sizes to enlarge these nodes' losses---making these nodes easier to be misclassified with graph perturbations. 
In contrast, we should put smaller weights on nodes with larger certified perturbation sizes, in order to save the usage of the perturbation budget. Formally, we 
assign the node weight 
such that $w(u) \sim 1/{K(\underline{p_{y_u}})}$. 
There are many ways to assign node weights satisfying the inverse relationship. 
In this paper, 
for instance, we 
propose to 
define node weights as 
\begin{align}
\label{eqn:weight}
w(u) = \frac{1}{1+\exp(a \cdot K(\underline{p_{y_u}}))}, 
\end{align}
where $a$ is a tunable hyperparameter. 
We can observe that the node weight is exponentially decreased as the node's certified perturbation size increases. Such a property ensures that the majority of perturbed edges are used for disrupting 
nodes with smaller certified perturbation sizes 
(See Figures~\ref{fig:distpwcr_evasion}) when performing the attack.

\subsection{Certified Robustness Inspired Attack Design}
\label{sec:attackdesign}
Based on the derived certified perturbation size and our certified robustness inspired loss, we now propose to generate graph perturbations against GNNs with both graph evasion and poisoning attacks. 

\noindent {\bf Certified robustness inspired graph evasion attacks to generate graph perturbations.} 
We can choose any graph 
evasion attack to GNNs as the base evasion attack. 
In particular, given the attack loss from any existing evasion attack, we only need to modify the loss by multiplying it with our certification perturbation sizes defined node weights. 
For instance, we can use the PGD attack~\cite{xu2019topology} as the base evasion attack.
We replace its attack loss by 
our certified robustness inspired loss $\mathcal{L}_{CR}$ in Equation~\ref{eq:cr_loss}.  
Then, we have our certified robustness inspired PGD (CR-PGD) evasion attack that iteratively generates 
graph perturbations as follows:  
\begin{align}
\label{eqn:crpgd}
\delta = \textrm{Proj}_{\mathbb{B}} ( \delta + \eta \cdot \nabla_\delta \mathcal{L}_{CR}(f_\theta, \mathbf{A} \oplus \delta,\mathcal{V}_{Te})),
\end{align}
where $\eta$ is the learning rate in PGD, $\mathbb{B} = \{ \delta: \mathbf{1}^T \delta \le \Delta, \delta \in [0,1]^{\mathcal{|V|} \times \mathcal{|V|}} \}$ is the allowable perturbation set, and 
\begin{small}
\begin{align}
\textrm{Proj}_{\mathbb{B}}(\mathbf{a}) = 
\begin{cases}
\Pi_{[0,1]}(\mathbf{a} - \mu \mathbf{1}), & \textrm{if } 
\mathbf{1}^T \Pi_{[0,1]}(\mathbf{a} - \mu \mathbf{1}) = \Delta, \\
\Pi_{[0,1]}(\mathbf{a}), & \textrm{if } \mathbf{1}^T \Pi_{[0,1]}(\mathbf{a}) \leq \Delta,
\end{cases}
\end{align}
\end{small}%
where $\mu>0$, $\Pi_{[0,1]}(x)=x$ if $x \in [0,1]$, 0 if $x<0$, and 1 if $x>1$. 
The final graph perturbation is used to perform the evasion attack.

\noindent {\bf Certified robustness inspired graph poisoning attacks to generate graph perturbations.} 
Likewise, we can choose any graph 
poisoning
attack to GNNs as the base poisoning attack. 
Given the bilevel loss
from any existing poisoning attack, we simply modify each loss
by multiplying it with our certified perturbation sizes' defined
node weights. 
Specifically, we have
\begin{align} 
& \max_{\delta} \mathcal{L}_{CR}(f_{\theta^*}, \mathbf{A} \oplus \delta, \mathcal{V}_{Tr}), \label{poison_trainloss_cr} \\
& \textrm{s.t. } 
\theta^* = \arg\min_\theta \mathcal{L}_{CR}(f_\theta, \mathbf{A} \oplus \delta, \mathcal{V}_{Tr}), \, ||\delta||_0 \le \Delta, 
\label{poison_trainloss_cons}
\end{align}
where $\mathcal{L}_{CR}(f_\theta, \mathbf{A} \oplus \delta, \mathcal{V}_{Tr}) = \sum_{v \in \mathcal{V}_{Tr}} w(v) \cdot \ell({ f_{\theta}(\mathbf{A} \oplus \delta, v) \, y_v})$. 
Then, solving Equation~\ref{poison_trainloss_cr} and
Equation~\ref{poison_trainloss_cons} 
produces the poisoning attack graph perturbations with our framework.

Algorithm~\ref{alg:cr_PGD} and Algorithm~\ref{alg:cr_minmax} in Appendix 
show two instances of applying our CR inspired attack framework to the PGD evasion attack and Minmax~\cite{xu2019topology} poisoning graph, respectively. 
To save time, we calculate nodes' certified perturbation sizes per ${INT}$ iterations. 
Then, comparing with PGD, the computational overhead of our CR-PGD is calculating the node's certified perturbation size with a set of $N$ sampled noises every $INT$ iterations, which only involves making predictions on $N$ noisy matrices
and is efficient. 
Note that the predictions are independent and can be also parallelized. 
Comparing with Minmax, the computational overhead of our CR-Minmax is to independently train (a small number of) $N$ models every $INT$ iterations, which can be implemented in parallel.

\section{Experiments}
\label{sec:eval}

\subsection{Setup}
\label{sec:exp_setup}

\vspace{-2mm}
\noindent {\bf Datasets and GNN models.} 
Following \cite{xu2019topology,zugner2019adversarial}, we evaluate our attacks on 
benchmark graph datasets, i.e., Cora, Citeseer~\cite{sen2008collective}, and BlogCataLogs~\cite{rossi2015network}. 
Table~\ref{dataset_stat} in Appendix 
shows basic statistics of these graphs. 
We choose Graph Convolutional Network (GCN)~\cite{kipf2017semi} as the targeted GNN model, also following \cite{xu2019topology,zugner2019adversarial}.

\begin{figure*}[!t]
\centering
\subfloat[{Cora}]{\includegraphics[width=0.24\textwidth]{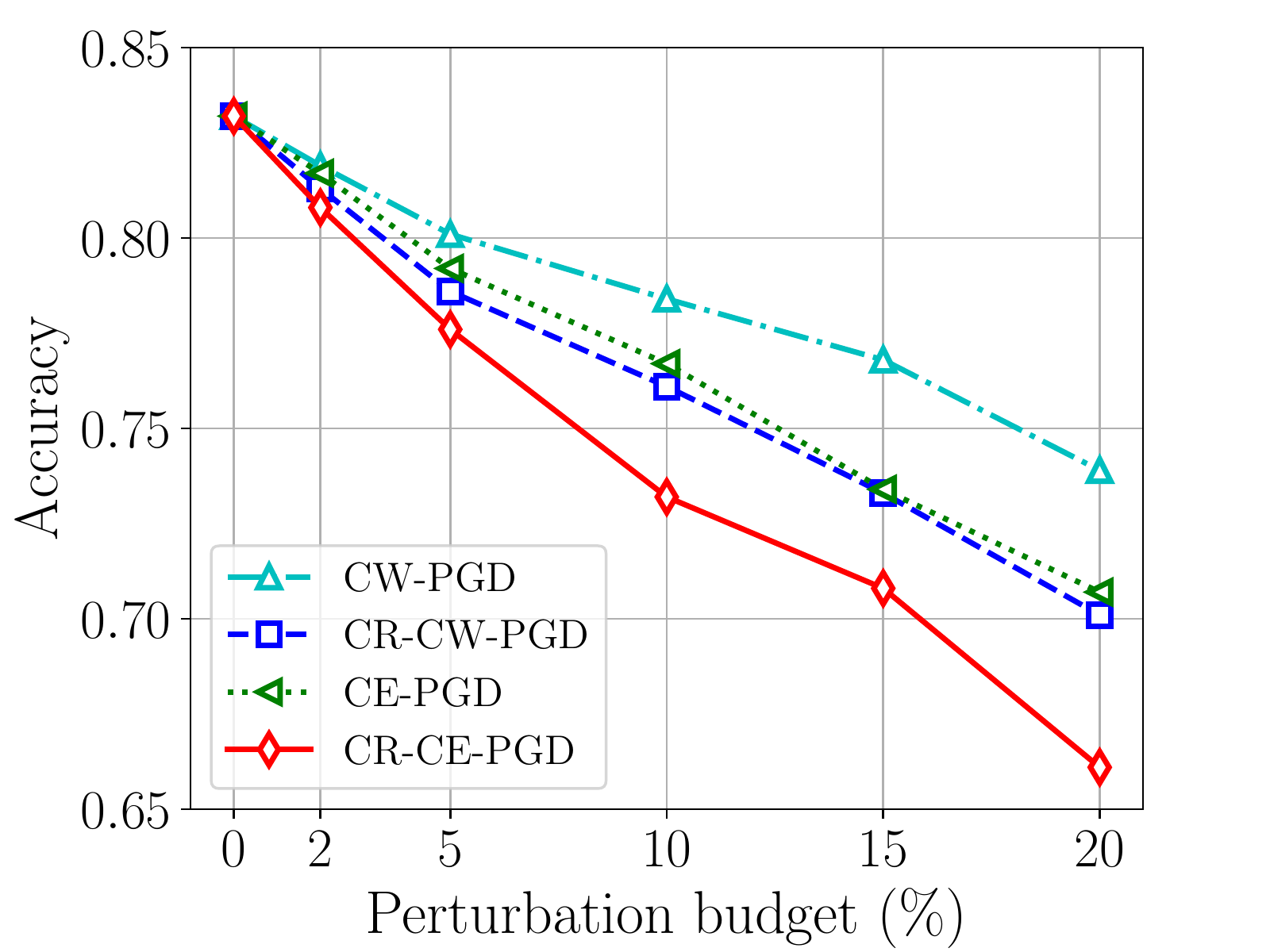}}
\subfloat[{Citeseer}]{\includegraphics[width=0.24\textwidth]{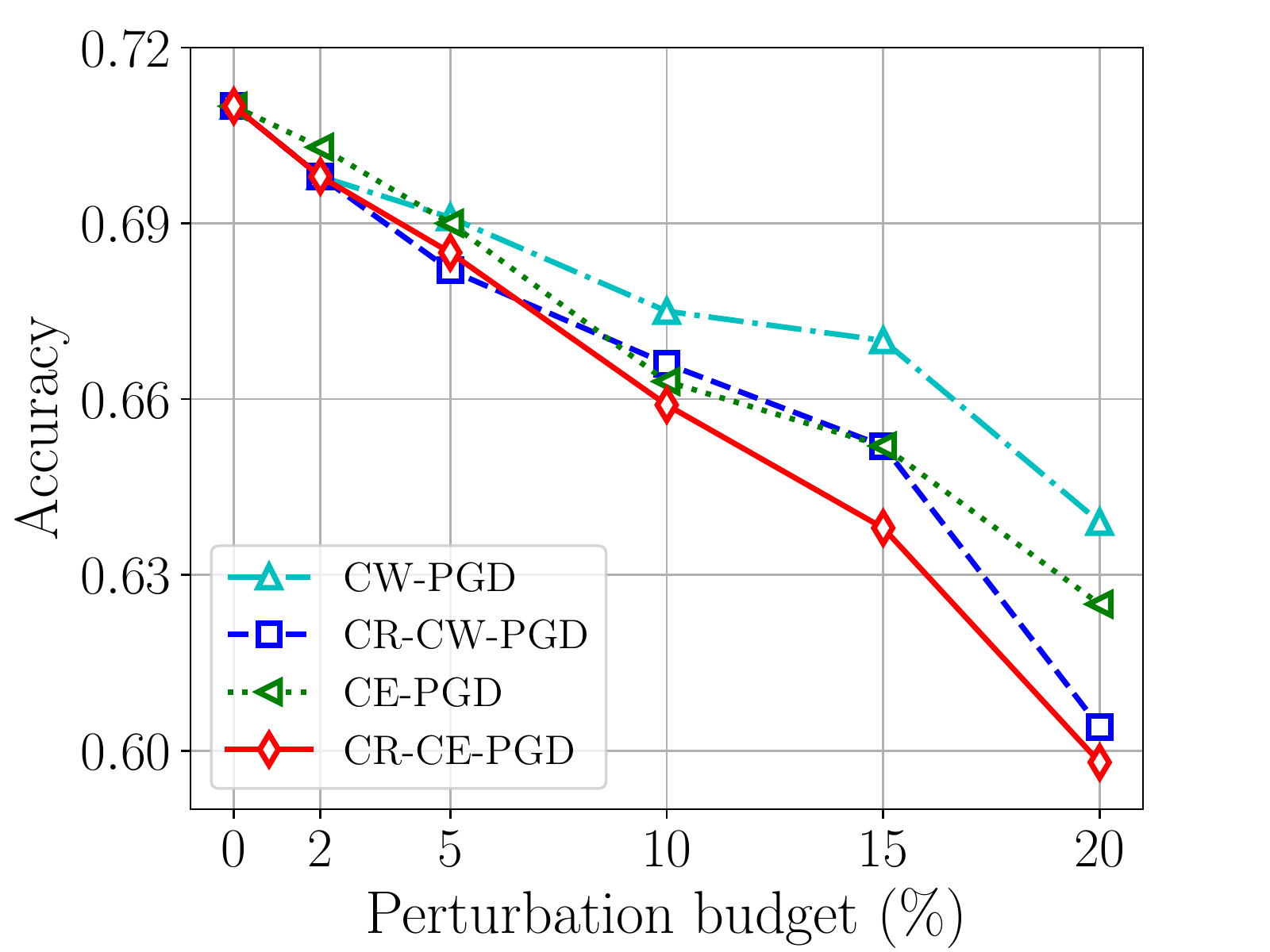}}
\subfloat[{BlogCataLog}]{\includegraphics[width=0.24\textwidth]{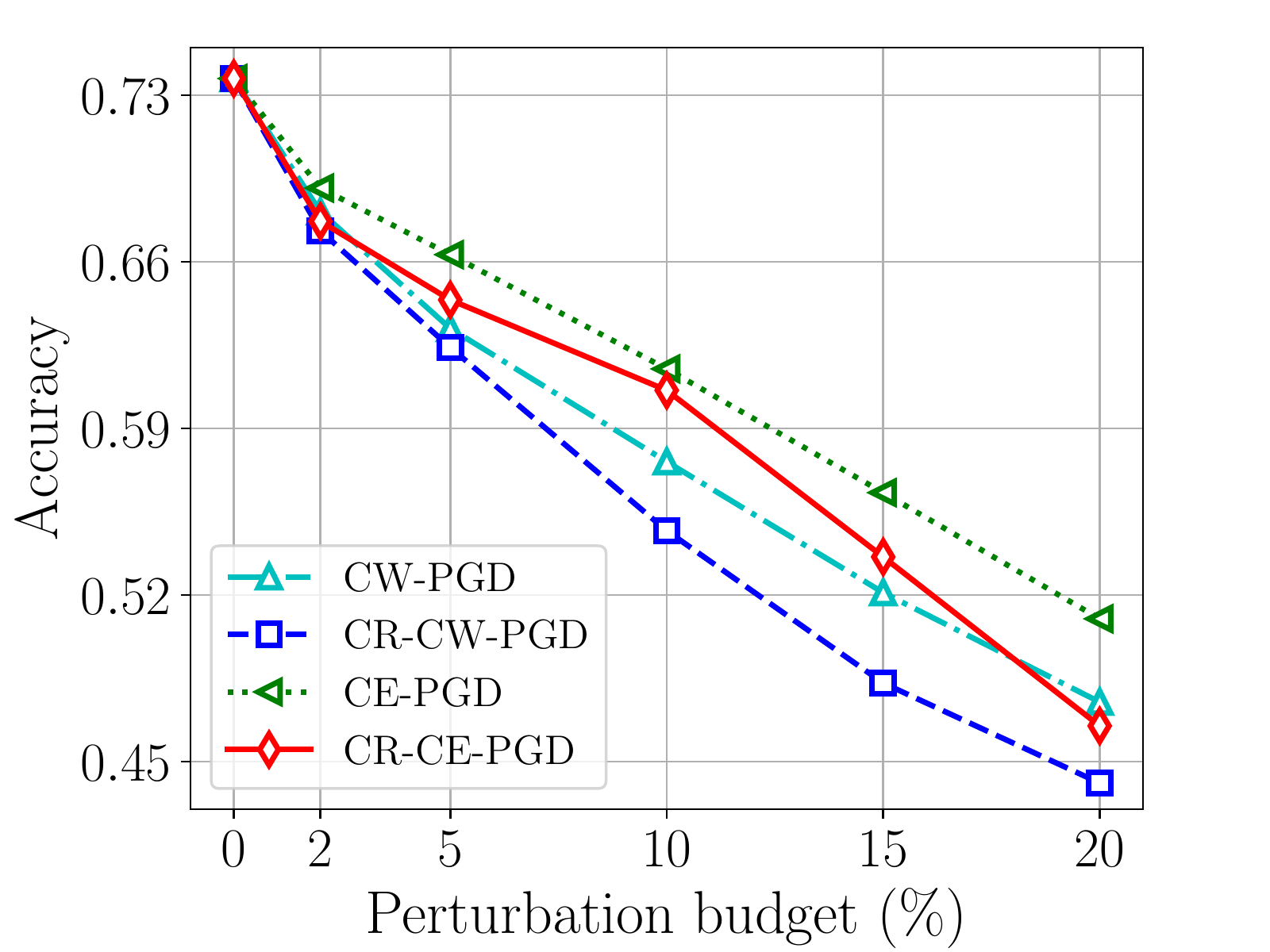}}
\vspace{-3mm}
\caption{Evasion attack accuracy vs. perturbation budget.} 
\label{fig:evasion_attack}
\vspace{-4mm}
\end{figure*}

\begin{figure*}[!t]
\centering
\subfloat[{  Cora}]{\includegraphics[width=0.24\textwidth]{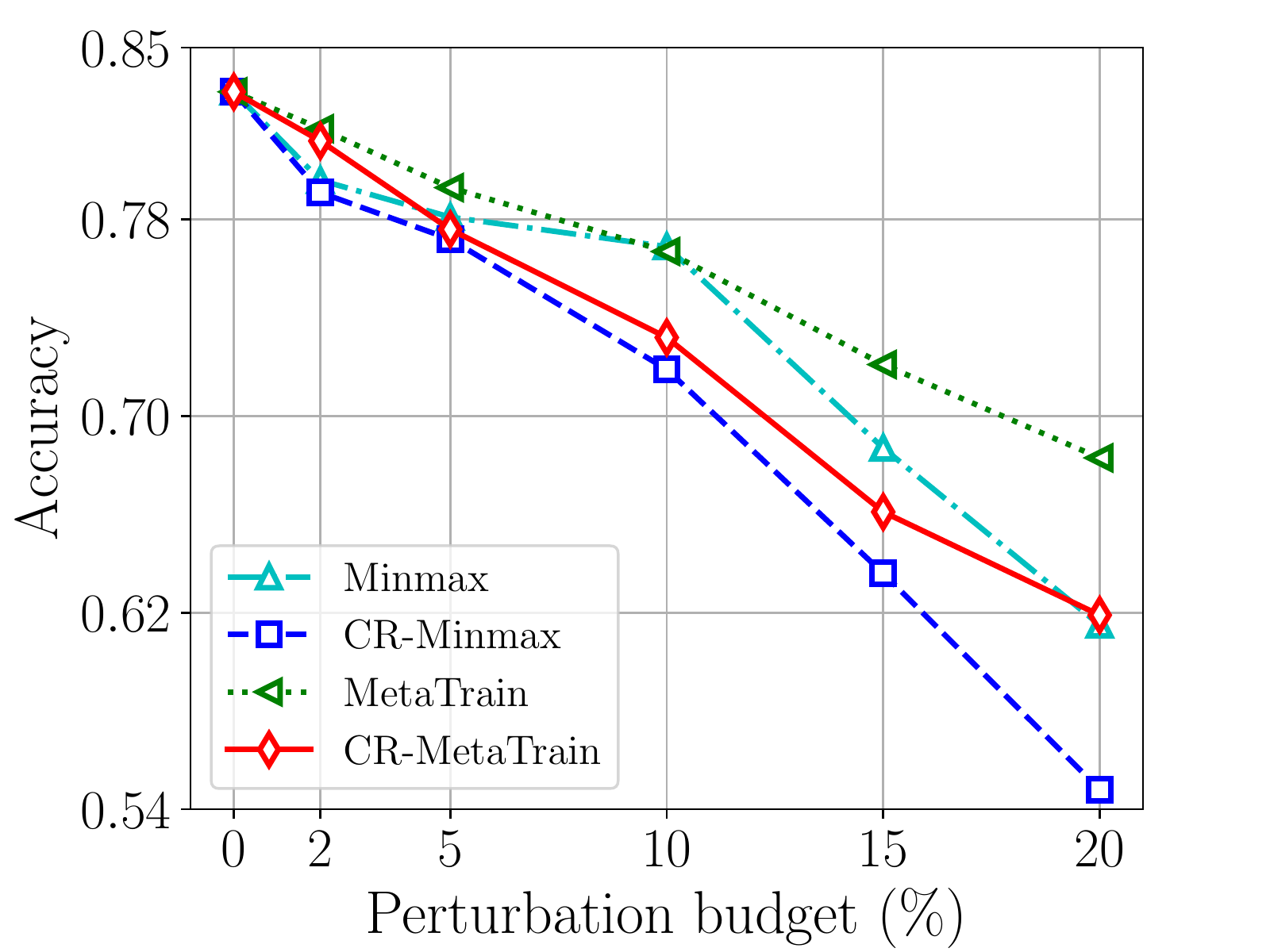}}
\subfloat[{  Citeseer}]{\includegraphics[width=0.24\textwidth]{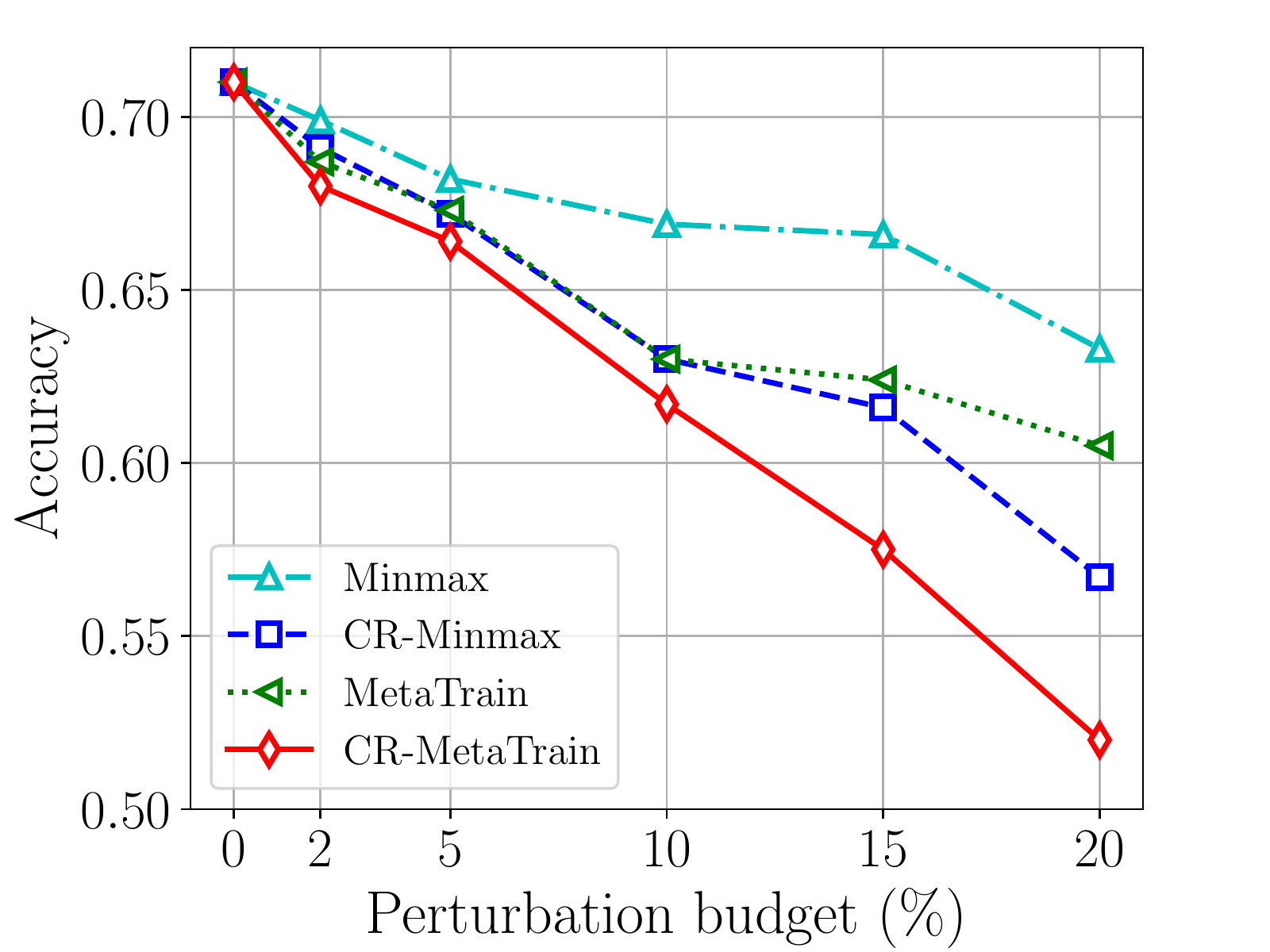}}
\subfloat[{  BlogCataLog}]{\includegraphics[width=0.24\textwidth]{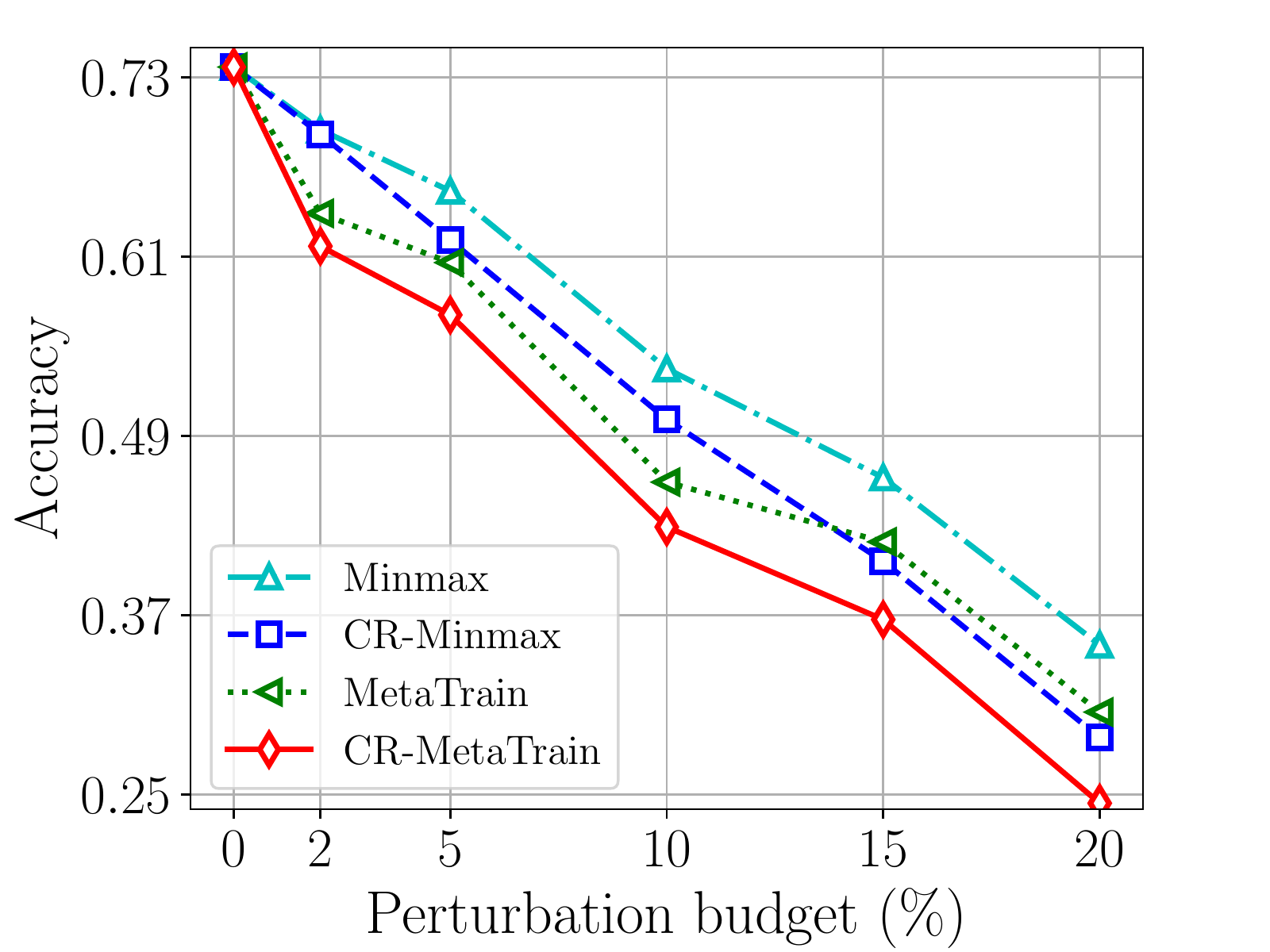}}
 \vspace{-3mm}
\caption{Poisoning attack accuracy vs. perturbation budget.} 
\label{fig:poisoning_attack}
\vspace{-4mm}
\end{figure*}

\vspace{-0.5mm}
\noindent {\bf Base attack methods.} 
For graph evasion attacks, we choose the PGD attack~\cite{xu2019topology}\footnote{\url{https://github.com/KaidiXu/GCN_ADV_Train}}  that uses the cross-entropy loss and CW loss~\cite{carlini2017towards} as the base attack methods, and denote the two attacks as CE-PGD and CW-PGD, respectively.
For graph poisoning attacks, we choose the Minmax attack~\cite{xu2019topology}
and MetaTrain attack~\cite{zugner2019adversarial}\footnote{\url{https://www.kdd.in.tum.de/gnn-meta-attack}} as the base attack methods.
We apply our CR inspired attack framework to these evasion and poisoning attacks and denote them as CR-CE-PGD, CR-CW-PGD, CR-Minmax, and CR-MetaTrain, respectively. 
All attacks are implemented in PyTorch and run on a Linux server with 96 core 3.0GHz
CPU, 768GB RAM, and 8 Nvidia A100 GPUs.

\noindent {\bf Training and testing.}
Following~\cite{zugner2019adversarial}, we split the datasets into 10\% training nodes, 10\% validation nodes, and 80\% testing nodes. The validation nodes are used to tune the hyperparameters, and the testing nodes are used to evaluate the attack performance. 
We repeat all attacks on 5 different splits of the training/evaluation/testing nodes and report the mean 
attack accuracy on testing nodes, i.e., fraction of testing nodes are misclassified after the attack.

\noindent {\bf Parameter settings.} 
Without otherwise mentioned, we set the perturbation budget $\Delta$ as 20\% of the total number of edges in a graph (before attack). We set the parameter $\beta=0.999$ in the noise distribution Equation~\ref{discretenoisedistribution}, 
the confidence level $1-\alpha=0.9$, 
the number of samples $N$ in Monte Carlo sampling to calculate node's certified perturbation size is set to be 200 and 20 in evasion attacks and poisoning attacks, respectively, 
and $a=1$ in Equation~\ref{eqn:weight}. 
The number of iterations $T$ is 100 and 10, and 
the interval is set to be $INT=10$ and $INT=2$ in evasion attacks 
and poisoning attacks, respectively. 
The other hyperparameters in CE-PGD, CW-PGD, Minmax, and MetaTrain are selected based on their source code, and we set equal values in our CR inspired attack counterparts. 
We also study the impact of the important hyperparameters that could affect our attack performance: $\Delta$, $N$, $1-\alpha$, $\beta$, and $a$. When studying the impact of a hyperparameter, we fix the other hyperparameters to be their default values.

\begin{figure*}[!t]
\centering
\subfloat[{  CW-PGD vs. CR-CW-PGD}]{\includegraphics[width=0.24\textwidth]{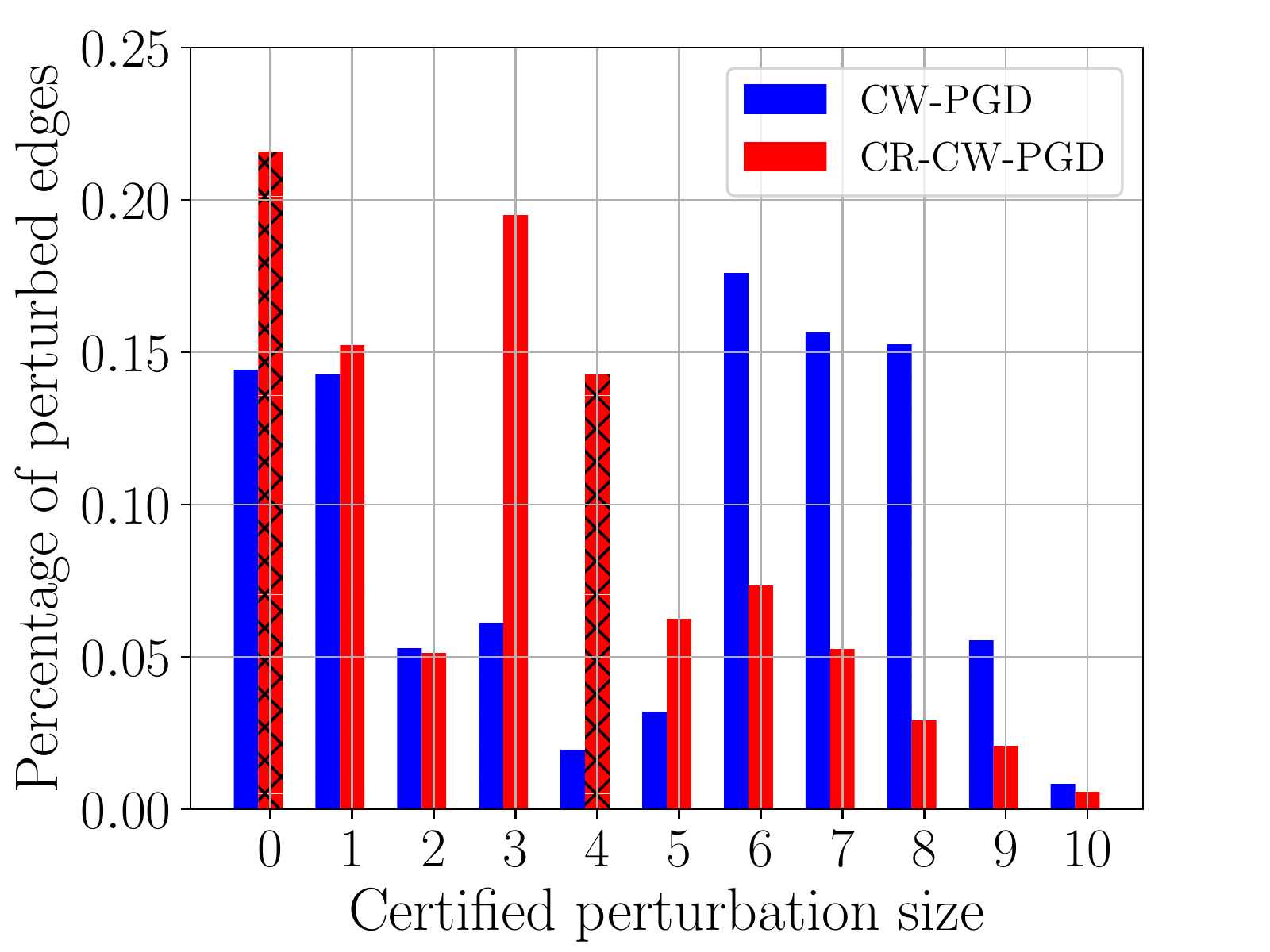}}
\subfloat[{  CE-PGD vs. CR-CE-PGD}]{\includegraphics[width=0.24\textwidth]{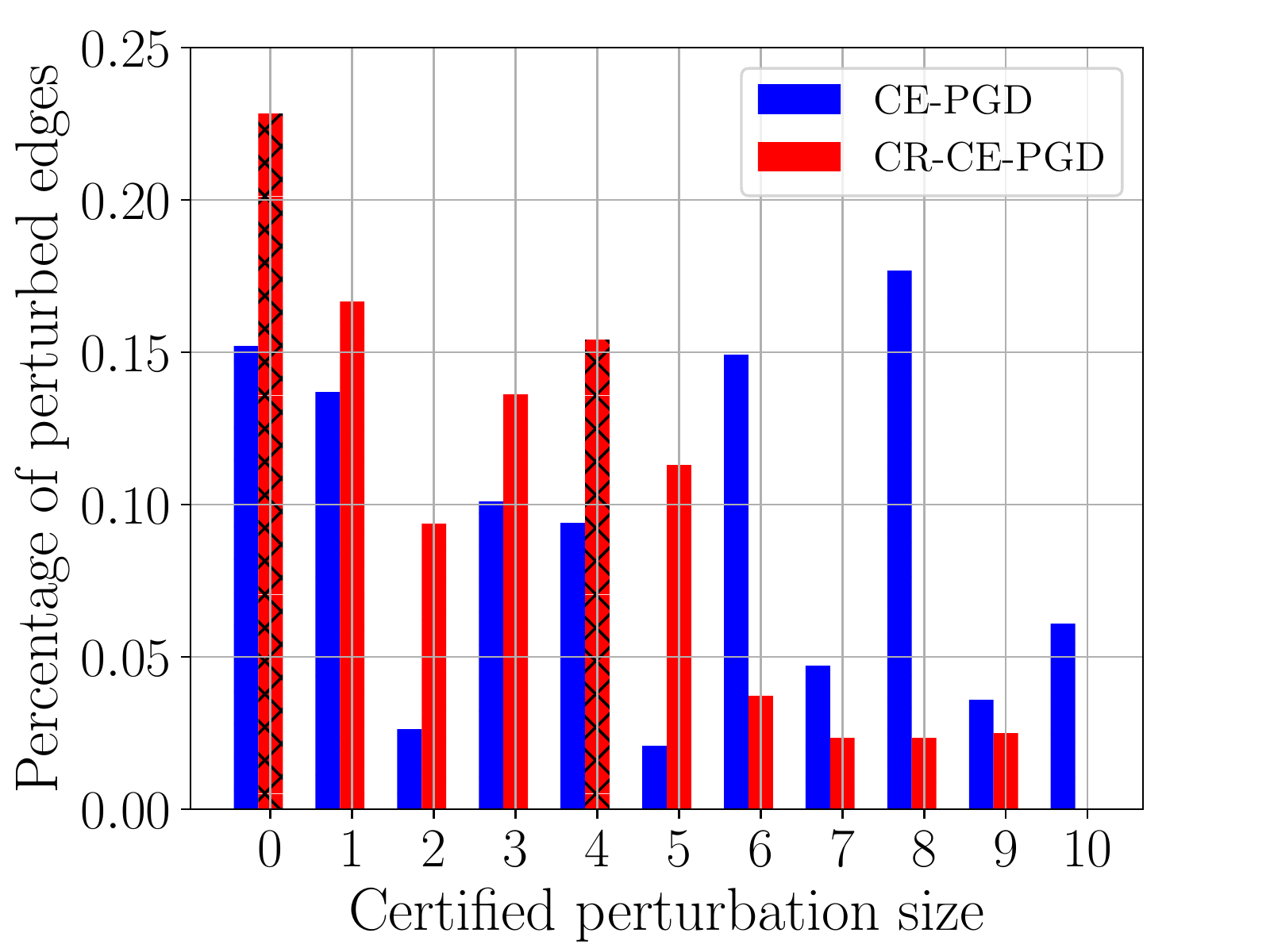}}
\subfloat[{  Minmax vs. CR-Minmax}]{\includegraphics[width=0.24\textwidth]{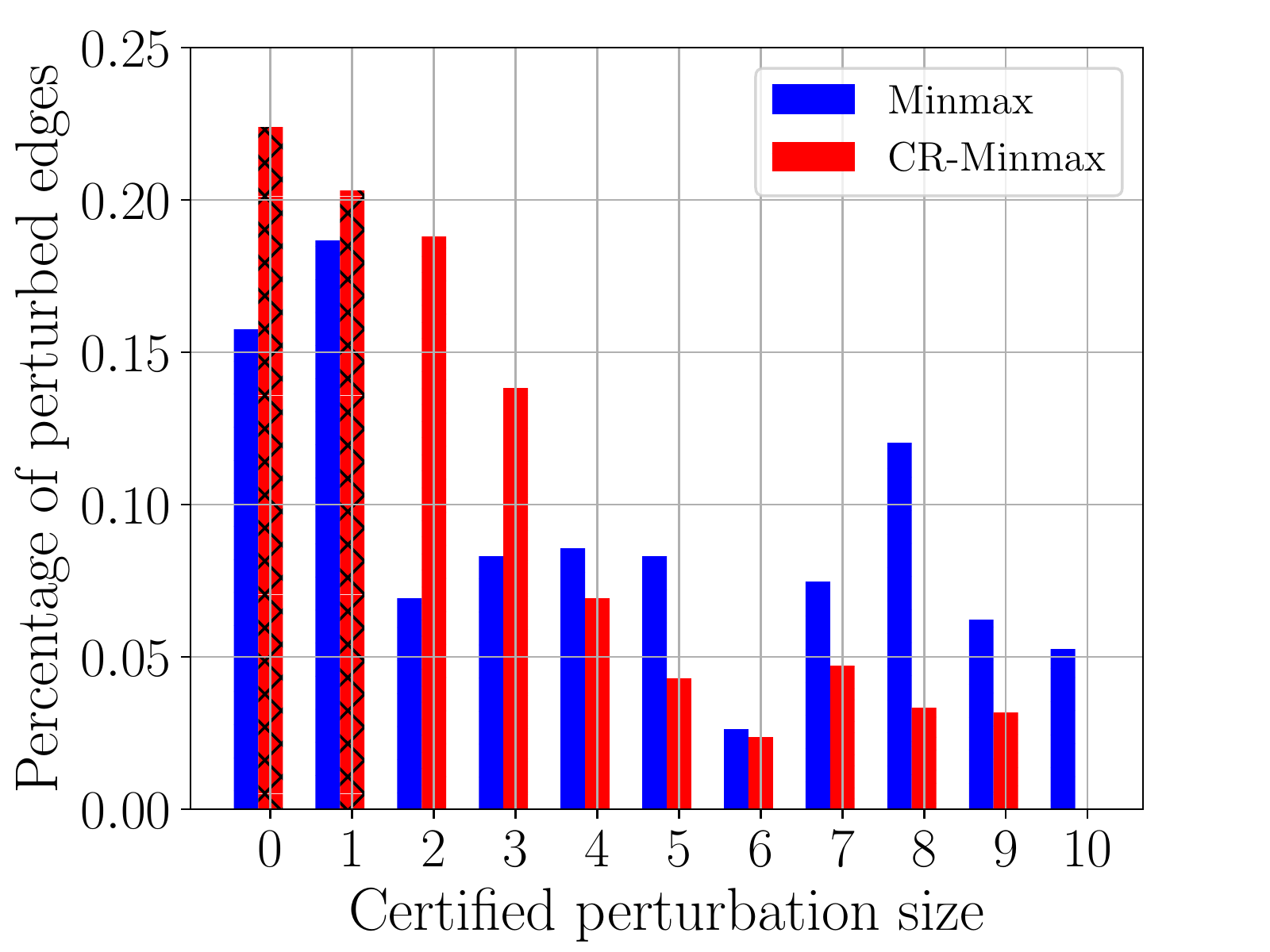}}
\subfloat[{  MetaTrain vs. CR-MetaTrain}]{\includegraphics[width=0.24\textwidth]{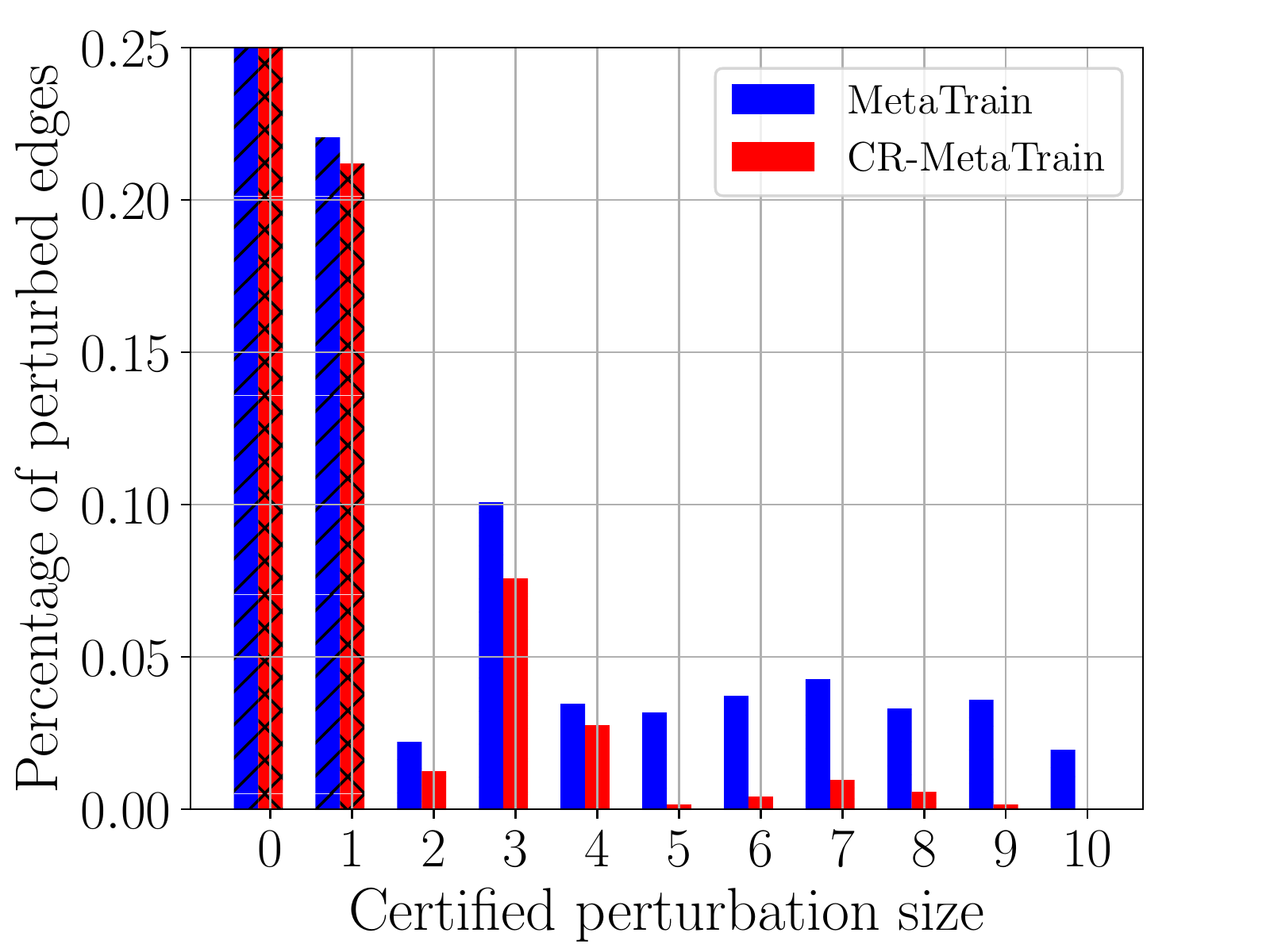}}
 \vspace{-3mm}
\caption{{Distribution of the perturbed edges vs. node's certified perturbation size on Citeseer.}} 
\label{fig:distpwcr_evasion}
\vspace{-6mm}
\end{figure*}

\subsection{Attack Results}
\vspace{-2mm}
\noindent {\bf Our attack framework is effective.} 
Figure~\ref{fig:evasion_attack} and Figure~\ref{fig:poisoning_attack} show the 
evasion attack accuracy and poisoning attack accuracy of the base attacks and those with our attack framework vs. perturbation budget, respectively. 
We can observe that \emph{our certified robustness inspired attack framework can enhance the base attack performance in all datasets.} For instance, when attacking GCN on Cora and the perturbation ratio is $20\%$, our CR-CE-PGD and CR-CW-PGD  have a relative $7.0\%$ and $5.6\%$ gain over the CE-PGD and CW-PGD evasion attacks. 
Moreover, CR-Minmax and CR-MetaTrain have a relative $12.2\%$ and $10.3\%$ gain over the Minmax and MetaTrain poisoning attacks. These results demonstrate that the node's certified robustness can indeed guide our attack framework to find the more vulnerable region in the graph to be perturbed, which 
helps to better allocate the perturbation budget, and thus 
makes the base attacks with our attack framework misclassify more nodes.

To further understand 
the effectiveness of our framework, we visualize the distribution of the perturbed edges 
vs. node's certified perturbation size. 
Specifically,  we first obtain the perturbed edges via the base attacks and our CR inspired attacks, and calculate testing/training nodes' certified perturbation sizes for evasion/poisoning attacks, respectively. 
Then we plot the distribution of the perturbed edges vs node's certified perturbation size. Specifically, if a perturbed edge is connected with a testing/training node in the evasion/poisoning attack, then we map this perturbed edge to this node's certified perturbation size. 
Our intuition is that a perturbed edge affects its connected node the most. Figure~\ref{fig:distpwcr_evasion} shows the results 
on Citeseer (We observe that the conclusions on the other datasets are similar).  
We can see that a majority number of the perturbed edges connect with testing/training nodes that have relatively smaller certified perturbation sizes in our CR inspired attacks. 
In contrast, a significant number of the perturbed edges in the base attacks connect with nodes with relatively larger certified perturbation sizes. 
Hence, under a fixed perturbation budget, our attacks can 
misclassify more nodes.

\noindent {\bf Comparing with other weight design strategies.}
Recall that our weight design is based on node's certified robustness:
nodes less provably robust to graph perturbations are  assigned larger weights, in order to enlarge these node attack losses. 
Here, we consider three other possible strategies to design node weights that aim to \emph{empirically} capture this property: 1) {\bf Random}, where we uniformly assign node weights between $[0, 1]$ at random; 
2) {\bf Node degree}, where a node with a smaller degree might be less robust to graph perturbations, and we set a larger weight. 
Following our weight design, we set $w_\textrm{deg}(u) = \frac{1}{1+\exp(a \cdot \textrm{deg}(u))}$;  
3) {\bf Node centrality}~\cite{newman2018networks}, where a node with a smaller centrality might be less robust to graph perturbations, and we set a larger weight. Similarly, we set $w_\textrm{cen}(u) = \frac{1}{1+\exp(a \cdot \textrm{cen}(u))}$. As a baseline, we also consider no node weights. 

Table~\ref{tbl:weight_design} shows the attack results by applying these weight design strategies to the existing graph evasion and poisoning attacks. We have the following observations: 1) {\bf Random} obtains the attack performance
even worse than {\bf No weight}'s. This indicates an inappropriate weight design can be harmful to the attack. 
2) Node {\bf Degree} and {\bf Centrality} perform slightly better than {\bf No weight}. One possible reason is that nodes with larger degree and centrality are empirically more robust to perturbations, which are also observed in previous works, e.g., \cite{zugner2018adversarial,wang2019attacking}. 
3) {Our} weight design strategy performs the best. This is because our weight design \emph{intrinsically}  captures nodes' certified robustness and thus yields  more effective attacks.

\begin{table}[!t]\renewcommand{\arraystretch}{0.8}
\scriptsize
\caption{Attack performance with different weight design.}
\vspace{-6mm}
\addtolength{\tabcolsep}{-2pt}
\center
\begin{tabular}{|c|c|c|c|c|c|}
\hline
          {\bf Dataset}        & {\bf Method} & {\bf CW-PGD} & {\bf CE-PGD} & {\bf Minmax} & {\bf MetaTrain} \\ \hline
\multirow{4}{*}{\bf Cora} 
& {\bf No weight} & 0.74 & 0.71 & 0.62 & 0.68 \\ \cline{2-6} 
& {\bf Random} & 0.77 & 0.75 & 0.65 & 0.72 \\ \cline{2-6} 
& {\bf Degree} & 0.72 & 0.70  & 0.61 & 0.66 \\ \cline{2-6} 
& {\bf Centrality} & 0.73 & 0.70  & 0.60 & 0.66 \\ \cline{2-6} 
& {\bf Ours} & {\bf 0.70} & {\bf 0.66}  & {\bf 0.55} & {\bf 0.62} \\ \hline \hline 
                  
\multirow{3}{*}{\bf Citeseer} 
& {\bf No weight} & 0.64 & 0.63 & 0.63 & 0.61 \\ \cline{2-6} 
& {\bf Random} & 0.66 & 0.66 & 0.68 & 0.64 \\ \cline{2-6} 
& {\bf Degree}  & 0.64 & 0.61 & 0.60 & 0.59 \\ \cline{2-6}
& {\bf Centrality}  & 0.64 & 0.62 & 0.60 & 0.58 \\ \cline{2-6}
& {\bf Ours} & {\bf 0.60} & {\bf 0.60} & {\bf 0.57} & {\bf 0.52} \\  \hline \hline 
               
\multirow{3}{*}{\bf B.C.Log} 
& {\bf No weight} & 0.48 & 0.51 & 0.35 & 0.31 \\ \cline{2-6} 
& {\bf Random} & 0.54 & 0.55 & 0.40 & 0.35 \\ \cline{2-6} 
   & {\bf Degree}  & 0.46 & 0.50 & 0.32 & 0.28 \\ \cline{2-6}
      & {\bf Centrality}  & 0.47 & 0.49 & 0.32 & 0.27 \\ \cline{2-6}
& {\bf Ours} & {\bf 0.44} & {\bf 0.46}  &  {\bf 0.29} & {\bf 0.24} \\ \hline 
\end{tabular}
\label{tbl:weight_design}
\vspace{-4mm}
\end{table}

\begin{table}[!t]\renewcommand{\arraystretch}{0.8}
\scriptsize
\caption{Attack performance with different $a$.}
\addtolength{\tabcolsep}{-4pt}
\center
\vspace{-6mm}
\begin{tabular}{|c|c|c|c|c|c|}
\hline
          {\bf Dataset}        & {$a$} & {\bf CR-CWPGD} & {\bf CR-CEPGD} & {\bf CR-Minmax} & {\bf CR-MetaTrain} \\ \hline
\multirow{3}{*}{\bf Cora} & {$0.5$} & 0.70 & 0.67 & 0.55 & 0.62 \\ \cline{2-6} 
                  & {$ 1$} & 0.70 & 0.66  & 0.55 & 0.62 \\ \cline{2-6} 
                  & {$ 2$} & 0.70 & 0.66  & 0.54 & 0.62 \\ \hline \hline
                  
\multirow{3}{*}{\bf Citeseer} & {$ 0.5$} & 0.60 & 0.61 & 0.58 & 0.54 \\ \cline{2-6} 
                  & {$ 1$} & 0.60 & 0.60 & 0.57 & 0.52 \\ \cline{2-6} 
                  & {$ 2$} & 0.60 & 0.59 & 0.57 & 0.53 \\ \hline \hline
                  
\multirow{3}{*}{\bf B.C.Log} & {$ 0.5$} & 0.44 & 0.47 & 0.31 & 0.25 \\ \cline{2-6} 
                  & {$ 1$} & 0.44 & 0.46  &  0.29 & 0.24 \\ \cline{2-6}
                  & {$ 2$} & 0.44 & 0.46 & 0.29 & 0.24 \\ \hline
\end{tabular}
\label{tbl:weight_a}
\vspace{-6mm}
\end{table}

\begin{figure*}[!t]
\centering
\subfloat[Noise parameter $\beta$]{\includegraphics[width=0.24\textwidth]{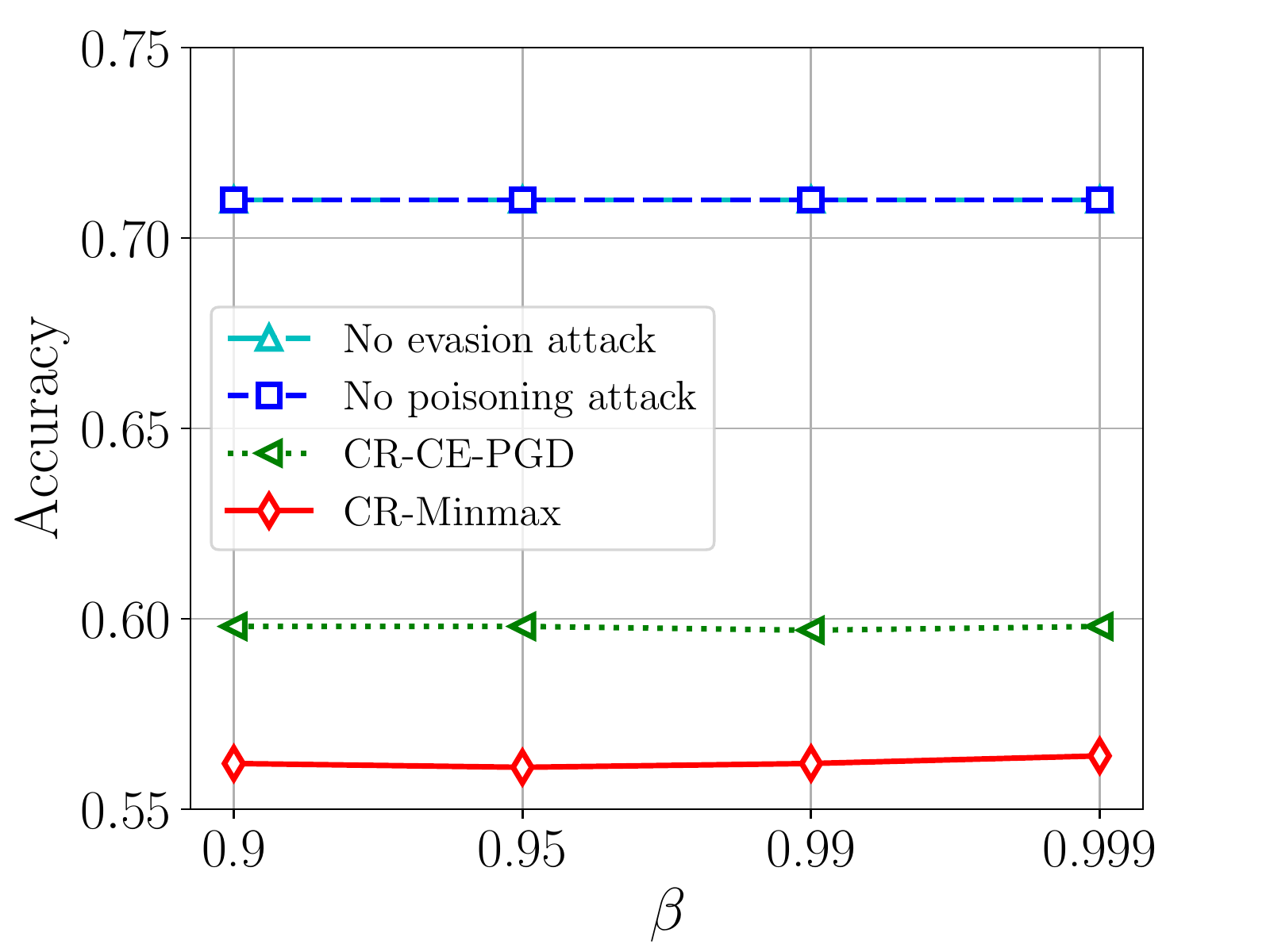}}
\subfloat[Confidence level $1-\alpha$]{\includegraphics[width=0.24\textwidth]{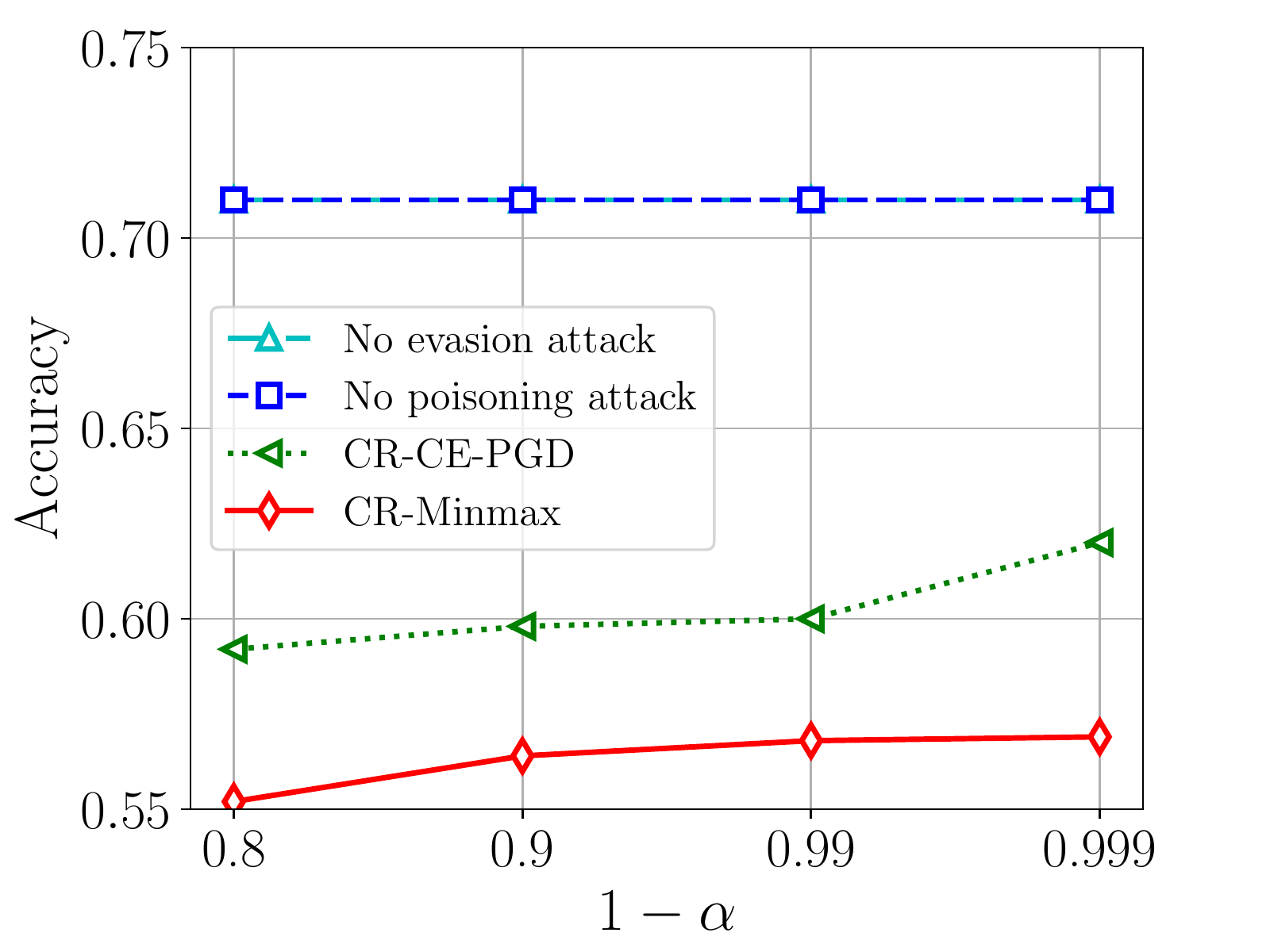}}
\subfloat[Number of samples
$N$]{\includegraphics[width=0.235\textwidth]{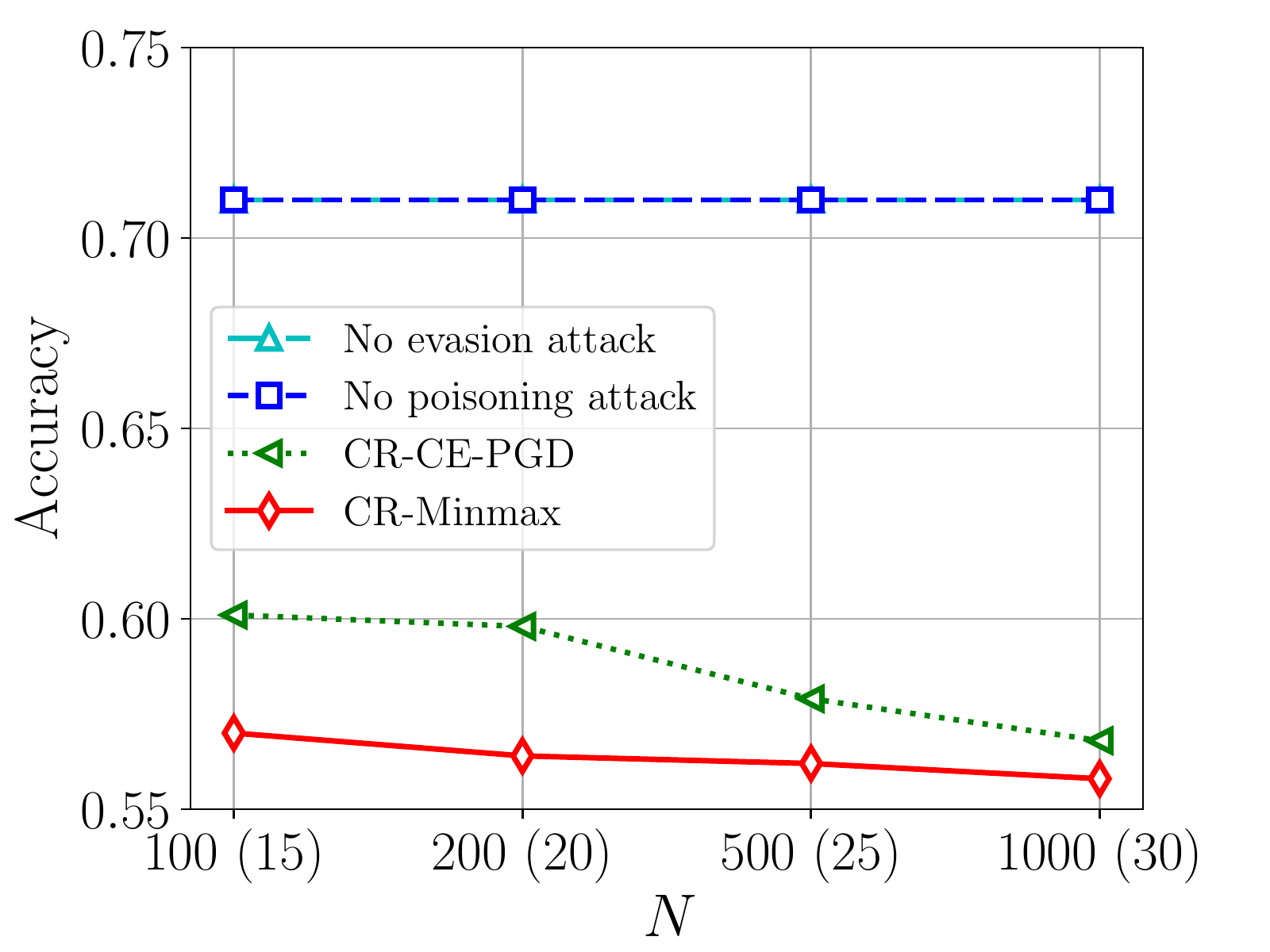}}
 \vspace{-3mm}
 \subfloat[Running time vs. 
$N$]{\includegraphics[width=0.235\textwidth]{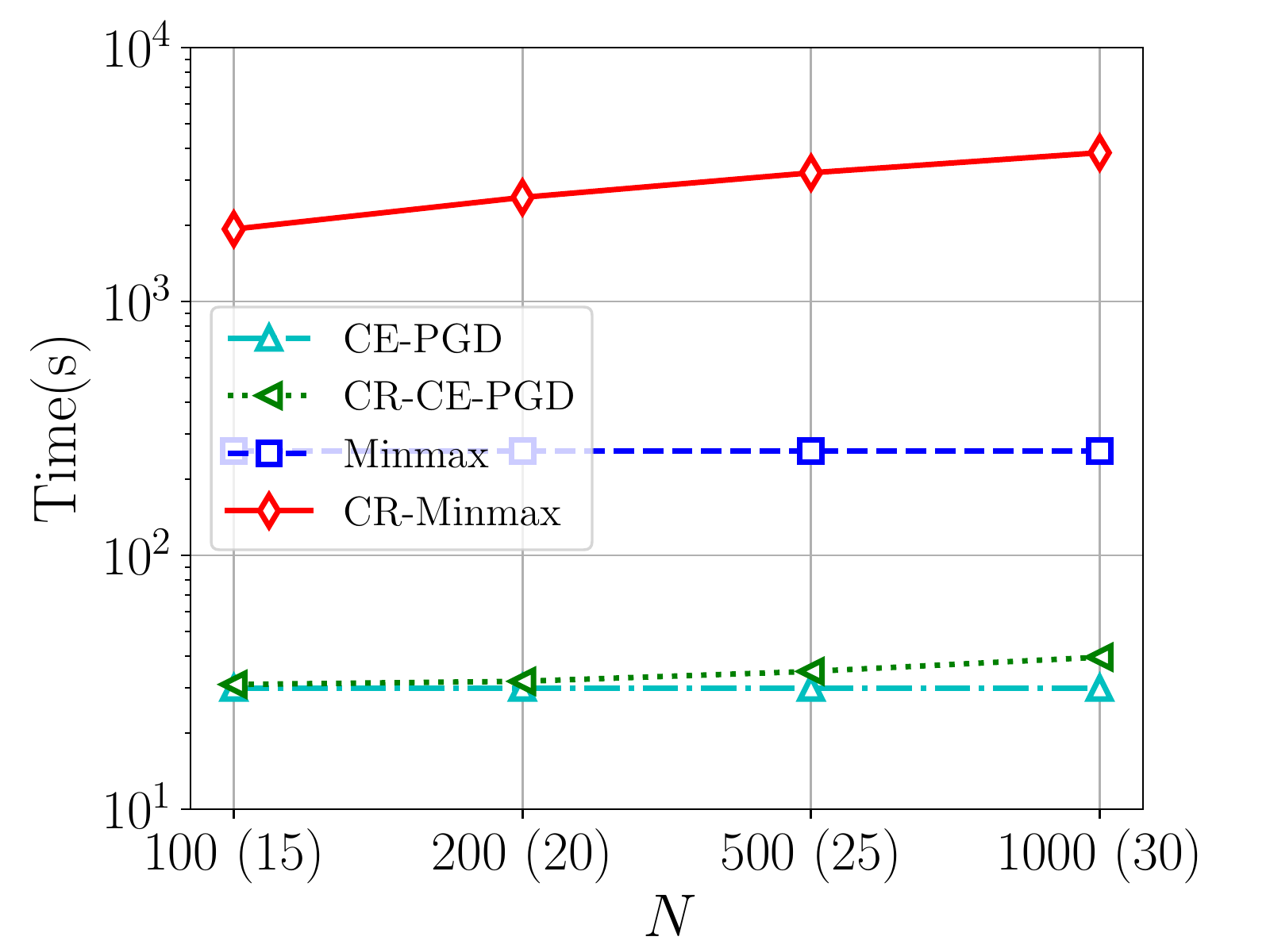}}
\caption{Impact of (a) $\beta$, (b) $1-\alpha$, (c) $N$ (\# in bracket in x-axis is for poisoning attacks), and (d) running time vs. $N$ on Citeseer. Note that (c): ``No evasion attack" and ``No poisoning attack" curves are overlapped; (d) $INT$=10 (2) for our evasion (poisoning) attacks.} 
\label{fig:impact_M_noise}
\vspace{-6mm}
\end{figure*}

\noindent {\bf Ablation study.}
In this experiment, we study the impact of 
hyperparameters:  
$\beta$ in Equation~\ref{discretenoisedistribution}, confidence level $1-\alpha$ in Equation~\ref{eqn:lowbound}, and $N$ in 
Equation~\ref{eqn:lowbound}, $a$ in Equation~\ref{eqn:weight}, as well the running time vs. $N$.
Figure~\ref{fig:impact_M_noise} shows the results of $\beta$, $1-\alpha$, and $N$, and running time vs. $N$
on our attack.
We observe that: 1) Our attack is not sensitive to $\beta$; 2) 
Our attack slightly becomes worse as the confidence level $1-\alpha$ increases. Such an observation can guide an attacker to set a relatively small $\beta$ in practice.  
3) Our attack becomes better as $N$ increases, but already works well with a relatively smaller $N$. From this observation, an attacker can choose a small $N$ in practice to save the time and cost when performing the attack. 
4) Running time does not increase too much with $N$ on the evasion attacks and is linear to $N$ on poisoning attacks, consistent with our analysis in Section~\ref{sec:attackdesign}. 

Table~\ref{tbl:weight_a} shows the impact of  $a$. We see the performances are stable across different $a$. This is  
largely because our weight design already ensures the node weight is inversely and exponentially to the node's
certified perturbation size.

\section{Discussion}

\vspace{-2mm}
\noindent {\bf Evaluations on other GNNs.}
We mainly follow existing attacks~\cite{xu2019topology,zugner2019adversarial}, which only  evaluate GCN. Here, we also test SGC~\cite{wu2019simplifying} on Cora and results show our CR-based GNN attacks also have a 6\%-12\% gain over the base attacks. This validates our strategy is generic to design better attacks. 

\noindent {\bf Transferability between different GNNs.} 
We evaluate the transferability of the graph perturbations generated by our 4 CR-based attacks on GCN to SGC on Cora, when the attack budget is 15. 
Accuracy on SGC under the 4 attacks are: 73\%, 76\%, 66\%, and 67\%, while accuracy on GCN are 71\%, 73\%, 63\%, and 65\%, respectively. This implies a promising transferability between GCN and SGC.

\noindent {\bf Defenses against our attacks.} 
Almost all existing empirical defenses~\cite{zhu2019robust,wu2019adversarial,entezari2020all,zhang2020gnnguard,jin2020graph,geisler2021robustness,zhuang2022defending} are ineffective to adaptive attacks~\cite{mujkanovic2022defenses}.
We adopt adversarial training~\cite{madry2017towards}, which is the only known effective empirical defense. 
Specifically, we first generate graph
perturbations 
for target nodes via our attack and use the perturbed graph to retrain GNN with true node labels. The trained GNN is used for evaluation. 
We test on Cora and show this defense is effective to some extent, but has a nonnegligible utility loss. For instance, when budget=15, the accuracy under
the CR-CW-PGD (CR-CE-PGD) attack increases from 73\%
(71\%) to 76\% (73\%), but the normal accuracy reduces from
84\% to 73\% (72\%).

\vspace{-2mm}
\section{Related Work}
\label{sec:related}

\vspace{-2mm}
\noindent {\bf Attacking graph neural networks (GNNs).}
We classify the existing attacks to GNNs as {evasion attacks}~\cite{dai2018adversarial,zugner2018adversarial,xu2019topology,wu2019adversarial,ma2019attacking,ma2020towards,mu2021hard,wang2022bandits} and {poisoning attacks}~\cite{zugner2018adversarial,dai2018adversarial,zugner2019adversarial,xu2019topology,takahashi2019indirect,liu2019unified,sun2020adversarial,zhang2020backdoor}.  
E.g., 
 Xu et al.\cite{xu2019topology} 
proposed an untargeted PGD graph evasion attack to the GCN. The PGD attack leverages first-order optimization and generates discrete graph perturbations via convexly relaxing the binary graph structure, and obtains the state-of-the-art attack performance.
Regarding graph poisoning attacks, 
Zugner et al.\cite{zugner2019adversarial} proposed a graph poisoning attack, called Metattack, that perturbs the whole graph based on meta-learning. 
Our attack framework can be seamlessly plugged into these graph evasion and poisoning attacks and enhance their attack performance.

\noindent {\bf Attacking other graph-based methods. }
Besides attacking GNNs, other adversarial attacks against graph data include attacking graph-based clustering~\cite{chen2017practical}, graph-based collective classification~\cite{torkamani2013convex,wang2019attacking}, graph embedding~\cite{dai2018adversarialnet,chen2018fast,bojchevski2019adversarial,chang2020restricted}, community detection~\cite{li2020adversarial}, 
graph matching~\cite{zhang2020adversarial}, etc. 
For instance,  Chen et al.~\cite{chen2017practical} proposed a practical attack against spectral clustering, which is a well-known graph-based clustering method. 
Wang and Gong~\cite{wang2019attacking} designed an attack to the collective classification method, called linearized belief propagation, by modifying the graph structure.

\noindent {\bf Certified robustness and randomized smoothing.} 
Randomized smoothing~\cite{lecuyer2018certified,li2018second,cohen2019certified,zhai2020macer,levine2020robustness,yang2020randomized} was the first method to obtain certified robustness of large models and achieved state-of-the-art performance. For instance, Cohen et al.~\cite{cohen2019certified} leveraged the Neyman-Pearson Lemma~\cite{neyman1933ix} to obtain a tight $l_2$ certified robustness for randomized smoothing with Gaussian noise on normally trained 
image models. 
Salman et al.~\cite{salman2019provably} improved the certified robustness by combining the design of an adaptive attack against smoothed soft image classifiers and adversarial training on the attacked classifiers. 
\cite{jia2020certifiedcommunity}, \cite{wang2021certified}, and  \cite{bojchevski2020efficient} 
applied randomized smoothing in the graph domain and derived certified robustness
for community detection and node/graph classifications methods
against graph perturbations.  
In this paper, we use randomized smoothing
to design better attacks against GNNs.

\vspace{-2mm}
\section{Conclusion}
\label{sec:conclusion}
\vspace{-2mm}
We study graph evasion and poisoning attacks to GNNs
and propose a novel attack framework motivated by certified robustness. 
We are the first work that uses certified robustness for an attack purpose.
In particular, we first derive the node's certified perturbation size, by extending randomized smoothing from the classifier perspective to a general
function perspective. 
Based on it, we design {certified robustness} inspired node weights,
which can be seamlessly plugged into the existing graph perturbation 
attacks' loss and 
produce our certified robustness inspired attack loss and attack framework. 
Evaluations on multiple datasets demonstrate that existing attacks' performance can be significantly enhanced by applying our attack framework.

\vspace{+0.5mm}
\noindent{\bf Acknowledgments.} 
This work was supported by Wang's startup funding, the Cisco Research Award, and the National Science
Foundation under grant No. 2216926. Any opinions,
findings, and conclusions or recommendations expressed in this
material are those of the author(s) and do not necessarily reflect
the views of the funding agencies.

{
\bibliographystyle{ieee_fullname}
\bibliography{refs}
}

\newpage

\clearpage
\appendix

\begin{table}[!t]
\centering
\caption{Dataset statistics.}
\vspace{-3mm}
\footnotesize
\centering
\begin{tabular}{|c|c|c|c|c|c|c|} \hline 
{\bf Dataset}  & {\bf \#Nodes} & {\bf \#Edges} & {\bf \#Features} & {\bf \#Classes} \\ \hline
{\bf Cora}  &  {\bf 2,485} & {\bf 5,429} & {\bf 1,433}   & {\bf 7} \\ \hline
{\bf Citeseer}  &  {\bf 2,110} & {\bf 3,757} & {\bf 3,703}   & {\bf 6} \\ \hline
{\bf BlogCataLog} &  {\bf 5,196} & {\bf 343,486} & {\bf 8,189}   & {\bf 6}  \\ \hline
\end{tabular}
\label{dataset_stat}
\end{table}

\begin{algorithm}[h]
\small
\caption{Certified robustness inspired PGD (CR-PGD) graph evasion attack to GNNs}
\label{alg:cr_PGD}
\LinesNumbered
\KwIn{Node classifier $f$, graph $G(\mathbf{A})$, testing nodes $\mathcal{V}_{Te}$, perturbation budget $\Delta$, 
total iterations $T$, $\#$samples $N$, noise parameter $\beta$, confidence level $1-\alpha$, $a$,
interval $INT$.}
\KwOut{Adversarial graph perturbation $\delta^{(T)}$.}
Initialize: $t=0$; 
$\delta^{(0)} = 0$; \\
\While{$t<T$}
{
    {\bf // Stage 1: Obtaining the CR inspired loss}
    
    \eIf {t \textrm{mod} INT != 0}
    { 
    Reuse the node weights:
    $w^{(t)}(v)=w^{(t-1)}(v)$;
    }
    {
    Define the perturbed graph: $\mathbf{A}^{(t)} = \mathbf{A} \oplus \delta^{(t)}$;

    Sample $N$ noise matrices $\{\epsilon^j\}_{j=1}^N$ from the noise distribution Equation~\ref{discretenoisedistribution} with parameter $\beta$; 
    
    \For{each node $v \in \mathcal{V}_{Te}$}{
        Compute the frequency $N_{y_v}$ for label $y_v$: 
        $N_{y_v}=\sum_{j=1}^{N}\mathbb{I}({f}(\mathbf{A}^{(t)} \oplus \epsilon^j; v)=y_v)$; 
        
        Estimate the low bound probability $\underline{p_{y_v}}$ with confidence $1-\alpha$: $\underline{p_{y_v}} = B(\alpha; N_{y_v}, N-N_{y_v}+1)$; 
        
        Calculate the certified perturbation size $K(\underline{p_{y_v}})$ using $\underline{p_{y_v}}$ and algorithm in~\cite{wang2021certified}; 
        
        Assign a weight $w(v)$ to each node $v$: 
    $w^{(t)}(v)=\frac{1}{1+\exp(a \cdot K(\underline{p_{y_v}}))}$; 
    }
    }
    Define the certified robustness inspired test loss: 
    $\mathcal{L}_{CR}(f,\mathbf{A}^{(t)},\mathcal{V}_{Te}) = \sum_{v \in \mathcal{V}_{Te}} w^{(t)}(v) \ell(f(\mathbf{A}^{(t)}; v), y_v) $; 
    
    {\bf // Stage 2: Running the PGD attack with CR loss} 
    $\delta^{(t+1)}=\textrm{Proj}_{\mathbb{B}} (\delta^{(t)} + \eta \cdot \nabla_{\delta^{(t)}} \mathcal{L}_{CR}(f, \mathbf{A}^{(t)},\mathcal{V}_{Te})$; 
    
    Update $t=t+1$.
}
\Return{$\delta^{(T)}$}
\end{algorithm}

\begin{algorithm}[!t]
\small
\caption{Certified robustness inspired Minmax (CR-Minmax) graph poisoning attack to GNNs}
\label{alg:cr_minmax}
\LinesNumbered
\KwIn{GNN algorithm $\mathcal{A}$, Graph $G(\mathbf{A})$, training nodes $\mathcal{V}_{Tr}$, 
perturbation budget $\Delta$, number of samples $N$, noise parameter $\beta$, confidence level $1-\alpha$, $a$, interval $INT$.
}
\KwOut{Adversarial graph perturbation $\delta^{(T)}$.}
Initialize: $t=0$; $\delta^{(0)} = 0$;  
random/pretrained GNN model $\theta^{(0)}$; \\ 
\While{$t < T$} 
{
    {\bf // Stage 1: Obtaining the CR inspired loss}
    
    \eIf {t \textrm{mod} INT != 0}
    { 
    Reuse the node weights:
    $w^{(t)}(v)=w^{(t-1)}(v);$
    }
    
    {
        Define the perturbed graph: $\mathbf{A}^{(t)} = \mathbf{A} \oplus \delta^{(t)}$;
        
        Sample $N$ noise matrices $\{\epsilon^j\}_{j=1}^N$ from the noise distribution Equation~\ref{discretenoisedistribution} with parameter $\beta$; 
            
            Train $N$ node classifiers $\{\tilde{f}^n\}$ with  current perturbed graph $\mathbf{A}^{(t)}$ with the $N$ sampled noisy matrices $\{\epsilon^n\}$: $\tilde{f}^1 = \mathcal{A}(\mathbf{A}^{(t)} \oplus \epsilon^1, \mathcal{V}_{Tr}), \cdots, \tilde{f}^N=\mathcal{A}(\mathbf{A}^{(t)} \oplus \epsilon^N, \mathcal{V}_{Tr})$
            
            \For{each node $v \in \mathcal{V}_{Tr}$}{
                Compute the frequency $N_{y_v}$ for label $y_v$: 
                $N_{y_v}=\sum_{j=1}^{N}\mathbb{I}(\tilde{f}^j(\mathbf{A}^{(t)} \oplus \epsilon^j; v)=y_v)$; \\
                Estimate the low bound probability $\underline{p_{y_v}}$ with confidence $1-\alpha$: $\underline{p_{y_v}} = B(\alpha; N_{y_v}, N-N_{y_v}+1)$; \\
                Calculate the certified perturbation size $K(\underline{p_{y_v}})$ using $\underline{p_{y_v}}$ and algorithm in~\cite{wang2021certified}; \\
                Assign a weight $w(v)$ to each node $v$: 
            $w^{(t)}(v)=\frac{1}{1+\exp(a \cdot K(\underline{p_{y_v}}))}$;
            }
        }
    Define the certified robustness inspired training loss: 
        $\mathcal{L}_{CR}(f, \mathbf{A}^{(t)},\mathcal{V}_{Tr}) = \sum_{v \in \mathcal{V}_{Tr}} w_v^{(t)} \cdot \ell(f(\mathbf{A}^{(t)}; v), y_v) $; 
        
     {\bf // Stage 2: Running the Minmax attack with CR loss} \\
    {Step 1: Inner minimization over model parameter $\theta$:} 
    $\theta^{(t+1)} = \theta^{(t)} - \eta_1 \nabla_{\theta} \mathcal{L}_{CR}(f_{\theta^{(t)}}, {\bf A}^{(t)},  \mathcal{V}_{Tr}) $; 
    
    {Step 2: Outer maximization over graph perturbation $\delta$:} \\
    $\delta^{(t+1)} = \textrm{Proj}_{\mathbb{B}} ( \delta^{(t)} + \eta_2 \nabla_\delta \mathcal{L}_{CR}(f_{\theta^{(t+1)}}, \mathbf{A}^{(t)}, \mathcal{V}_{Tr}));$ \\
    Update $t = t+1$.
}
\Return{$\delta^{(T)}$} 
\end{algorithm}

\end{document}